\documentclass[twocolumn, tighten]{aastex631}

\usepackage{natbib}

\newcommand{\upenn}{Department of Physics \& Astronomy, University of Pennsylvania, 209 S 33rd St., Philadelphia, PA 19104, USA}
\newcommand{\cca}{Center for Computational Astrophysics, Flatiron Institute, 162 5th Ave, New York, NY 10010, USA}
\newcommand{\Msun}{\ensuremath{M_{\odot}}}
\graphicspath{{./}{figures/}}

\received{XXX}
\revised{YYY}
\accepted{ZZZ}

\shorttitle{Galaxy Progenitors of Stellar Streams}
\shortauthors{Panithanpaisal et al.}

\begin{document}

\title{The Galaxy Progenitors of Stellar Streams around Milky Way-mass Galaxies in the FIRE Cosmological Simulations}

\correspondingauthor{Nondh Panithanpaisal}
\email{nondh@sas.upenn.edu}

\author[0000-0001-5214-8822]{Nondh Panithanpaisal}
\affil{\upenn}

\author[0000-0003-3939-3297]{Robyn E. Sanderson}
\affil{\upenn}
\affil{\cca}

\author[0000-0003-0603-8942]{Andrew Wetzel}
\affil{Department of Physics \& Astronomy, University of California, Davis, CA 95616, USA}

\author[0000-0002-6993-0826]{Emily C. Cunningham}
\affiliation{\cca}

\author{Jeremy Bailin}
\affil{Department of Physics and Astronomy, University of Alabama, Box 870324, Tuscaloosa, AL, 35487, USA}

\author{Claude-Andr\'{e} Faucher-Gigu\`{e}re}
\affil{Department of Physics and Astronomy and CIERA, Northwestern University, 2145 Sheridan Road, Evanston, IL 60208, USA}



\begin{abstract}
Stellar streams record the accretion history of their host galaxy. We present a set of simulated streams from disrupted dwarf galaxies in 13 cosmological simulations of Milky Way (MW)-mass galaxies from the FIRE-2 suite at $z=0$, including 7 isolated Milky Way-mass systems and 6 hosts resembling the MW-M31 pair (full dataset at: \href{https://flathub.flatironinstitute.org/sapfire}{https://flathub.flatironinstitute.org/sapfire}). In total, we identify 106 simulated stellar streams, with no significant differences in the number of streams and masses of their progenitors between the isolated and paired environments. We resolve simulated streams with stellar masses ranging from $\sim5\times10^5$ up to $\sim10^{9}\Msun$, similar to the mass range between the Orphan and Sagittarius streams in the MW. We confirm that present-day simulated satellite galaxies are good proxies for stellar stream progenitors, with similar properties including their stellar mass function, velocity dispersion, [Fe/H] and [$\alpha$/H] evolution tracks, and orbital distribution with respect to the galactic disk plane.
Each progenitor's lifetime is marked by several important timescales: its infall, star-formation quenching, and stream-formation times. We show that the ordering of these timescales is different between progenitors with stellar masses higher and lower than $\sim2\times10^6 \Msun$. Finally, we show that the main factor controlling the rate of phase-mixing, and therefore fading, of tidal streams from satellite galaxies in MW-mass hosts is non-adiabatic evolution of the host potential. Other factors commonly used to predict phase-mixing timescales, such as progenitor mass and orbital circularity, show virtually no correlation with the number of dynamical times required for a stream to become phase-mixed.
\end{abstract}

\keywords{stellar stream, tidal stream, cosmological simulations}


\section{Introduction} \label{sec:intro}
Cosmological simulations predict that galaxies on the scale similar to the Milky Way (MW) contain significant number of gravitationally bound halos within their vicinity. However, the mass profiles and distribution of these halos are dependent on the detailed properties of the underlying dark matter model, especially its temperature. The scales where the primodial matter power spectrum is suppressed set the lower limit on their mass.  

In the cold dark matter (CDM) model, structures are formed in a hierarchical fashion. Small halos that are bound under gravitational influence of more massive halos, so called subhalos, merge to form smoother and more massive halos through tidal disruptions. CDM models, such as the weakly interacting massive particle (WIMP), suggest the halo lower mass limit as low as $10^{-11}M_\odot$ \citep{2009NJPh...11j5027B, PhysRevD.88.015027}. However, low-mass subhalos have little to no stellar content, rendering direct detections ineffective. Determining whether these dark subhalos exist, hence, serves as an important test to the CDM. The non-existence of these dark subhalos would favor alternative models of dark matter where there is a suppression in the primodial matter power spectrum at the scales of the classical dwarf galaxies, such as the warm dark matter, which predicts fewer low-mass subhalos (e.g. \cite{2003MNRAS.345.1285K})


Long cold stellar streams, which are tidally disrupted dwarf galaxies or globular clusters, can be used to detect subhalos in the MW. The stream is sensitive to heating by repeated encounters with low-mass ($\sim 10^{6}M_\odot$) dark subhalos \citep{2002MNRAS.332..915I, 2012ApJ...748...20C,2013ApJ...775...90C, 2016MNRAS.463..102E}. Simple simulations of the interaction of such streams with a subhalo even of mass $<10^7 M_\odot$ shows that it creates a discontinuity in the stream’s orbital energy distribution, which will later evolve into a gap or fluctuation in surface density of stars along the stream \citep{2011ApJ...731...58Y}. Numerous cold streams have been observed in the Milky Way of which some, such as GD-1 and Pal 5, exhibit features that could be due to interactions with subhalos \citep{2014ApJ...795...94B,2019ApJ...880...38B, 2012ApJ...760...75C, 2016ApJ...819....1I, 2016MNRAS.463.1759B}. These gaps, if really produced by stream-subhalo interaction, can possibly be modeled to estimate individual masses and mass profile of the interacting subhalo.

Formulating a robust stream-subhalo interaction model requires understanding dynamics of individual stars along the stream as well as dynamics of the entire stream within the main galaxy. These properties are directly related to the progenitor of each stream such as its initial mass, concentration, velocity dispersion and orbit. Many of these either have been assumed (such as inferring mass of the progenitor from velocity dispersion of the stream) or have only been studied in DM-only simulations. 

In this paper, we study the physical and dynamical properties of stellar streams and their dwarf galaxy progenitors in MW-like galaxies in a suite of fully cosmological-hydrodynamical simulations. Section 2 describes this suite, the FIRE-2 hydrodynamical simulations. Section 3 presents the criteria used to distinguish stellar streams from bound satellites and phase-mixed accreted structures in these simulations at the present day. In Section 4 we present the properties of the progenitor galaxies that form these streams. In Section 5 we examine the orbital characteristics of the progenitors as they form streams. In Section 6 we look at the time-evolution of the velocity dispersion in the streams and evaluate its usefulness as an indicator of the progenitor mass and/or the age of the stream.  In Section 7 we discuss how the stellar stream progenitors compare to satellite galaxies in the simulations that are still bound at the present day, and how they compare to observed satellite galaxies in the MW and M31. Section 8 summarizes our findings.

\section{Simulations} \label{sec:sims}

For this work, we use cosmological zoom-in baryonic simulations of MW-mass galaxies from the Feedback In Realistic Environments (FIRE) project\footnote{See the project website at: \href{http://fire.northwestern.edu}{http://fire.northwestern.edu}}. These simulations are run with \texttt{Gizmo} \citep{Hopkins2015}, which uses an optimized TREE+PM gravity solver and a Lagrangian mesh-free, finite-mass method for accurate hydrodynamics.
Star formation and stellar feedback are implemented using the state-of-the-art FIRE-2 physics model \citep{2018MNRAS.480..800H}, which adopts a ``bottom-up'' approach for modeling the dense multi-phase inter-stellar medium in galaxies and mono-age, mono-abundance stellar populations, and takes stellar feedback parameters directly from stellar evolution models like \texttt{STARBURST99} \citep{Leitherer1999}.

The FIRE-2 simulations produce MW-mass galaxies with many properties that are broadly consistent with observations of the MW, including: MW-like thin stellar disks with transient bars and spiral structure \citep{Ma2017, 2018MNRAS.480..800H, 2018arXiv180512199D, 2018arXiv180610564S}, a realistic distribution of massive GMCs \citep{2020MNRAS.492..488G, 2020MNRAS.497.3993B}, stellar halos from disrupted satellite galaxies \citep{2017arXiv171205808S}, and realistic populations of surviving satellite dwarf galaxies that do not suffer from the ``missing satellites'' or ``too-big-to-fail'' problems (\citealt{Wetzel2016}, \citealt{2018arXiv180604143G}, \citealt{Samuel2019}).
Part of the reason why satellite dwarf galaxies in FIRE-2 simulations agree well with observations of the MW (and M31) is that these simulations self-consistently form a massive central MW-like galaxy, which efficiently destroys satellite galaxies \textit{and low-mass subhalos} that orbit within the inner $\sim 30$ kpc via gravitational tidal forces \citep{2017MNRAS.471.1709G}. The tidal destruction leads to satellite radial distributions that agree with MW and M31, especially in the inner $\sim 150$ kpc of the halo, as well as reasonably accurate planes of satellite dwarf galaxies for some of the hosts \citep{2020MNRAS.491.1471S, 2020arXiv201008571S}.
The consistency of the phase-space volumes and orbits of the present-day satellites with observations, as well as the overall consistency of the stellar halo masses with observations, together support our expectation that a realistic population of streams should be created from tidally disrupted ``building blocks'' that are sufficiently resolved by the simulations. However, because of the particle resolution, stellar streams from globular clusters do not natively form in our simulations. In \S\ref{sec:proj}, we also explicitly compare these building blocks with the present day simulated satellites to determine whether this expectation is reasonable.

In this paper we study the tidal streams from dwarf galaxy progenitors around 13 MW-mass halos: 7 individual MW-mass halos and 3 Local-Group-like MW+M31 pairs. The isolated hosts are from the Latte suite, first presented in  \cite{2016ApJ...827L..23W}, while the paired hosts are from the ELVIS on FIRE suite, first presented in \cite{2019MNRAS.487.1380G}. All halos were simulated in $\Lambda$CDM cosmology at particle mass resolution of 3500--7100 $\Msun$ and spatial resolution of 1--4 pc for star/gas particles; 18,000--35,000 $\Msun$ and 40 pc for DM particles. Each simulation stores 600 snapshots from $z = 99$ to 0, providing post-processing time resolution of $\sim 20$ Myr---about 20 snapshots per orbit at 10 kpc, and more at larger galactocentric distances---allowing us to carry out dynamical analysis of the streams by post-processing.

The resolution of this suite of simulations allows both luminous and dark subhalos to be resolved well even near each MW-like galaxy, and follows the formation of tidal streams from dwarf galaxies down to slightly below the mass of the MW's ``classical`` dSphs: around $10^8\ \Msun$ in total mass or $10^6\ \Msun$ in stellar mass (at $z=0$). We will discuss the effects of resolution on our results throughout the paper. For now we note that given the steep mass function of stellar halo building blocks, this lower limit in stellar mass for resolved building blocks corresponds to more than 99\% of the total accreted mass in the stellar halo for a given simulation \citep{2015MNRAS.448L..77D, 2020MNRAS.497..747S}.

The dark matter particles in each snapshot of the finished simulations are processed with \texttt{Rockstar} \citep{2013ApJ...762..109B} to produce halo catalogs. These are connected in time using \texttt{consistent-trees} \citep{2013ApJ...763...18B} to form a merger tree. Once the merger tree is constructed, we make a preliminary assignment of star particles to each halo and subhalo identified by \texttt{Rockstar} in each snapshot, by selecting star particles within the halo's virial radius and within twice the halo circular velocity relative to the halo's center. This is a conservative criterion that excludes nearly all stream stars; we use it as a starting point for collecting the full stellar distribution as described in \S\ref{subsec:tracking}. The post-processing is done using \texttt{gizmo\_analysis} \citep{2020ascl.soft02015W} and \texttt{halo\_analysis} \citep{2020ascl.soft02014W}




\section{Selecting Stellar Streams in the Simulations}\label{sec:select} \label{sec:selection}

\begin{figure*}
\plotone{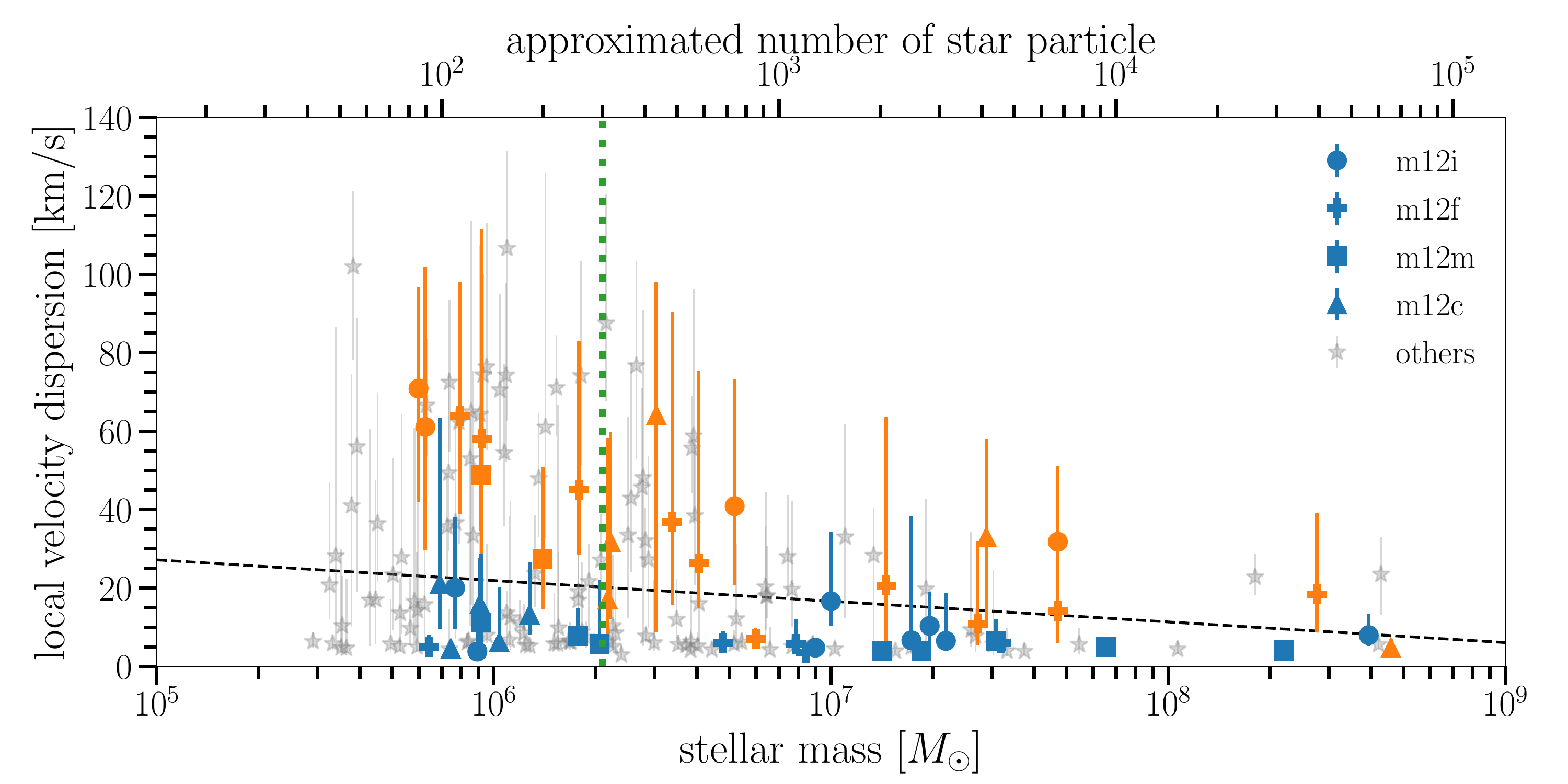}
\caption{Local velocity dispersions of stream candidates that meet criteria (i) and (ii) of \S\ref{sec:crit}. The marker is the median value and the error bar spans the $16^{\mathrm{th}}$ to $83^{\mathrm{rd}}$ percentiles of particles across the stream. The orange points are identified by eye as phase-mixed, while the blue points are identified by eye as streams. The gray points are all of the stream candidates from other simulations. Stream candidates above the local velocity dispersion threshold determined by the SVM (black dashed line) are classified as phase-mixed. The green vertical line marks the transition between using 7 nearest neighbors (to the left) and using 20 nearest neighbors (to the right) to estimate local velocity dispersions. \label{fig:dispersion_cut}}
\end{figure*}

\begin{figure*}
\plotone{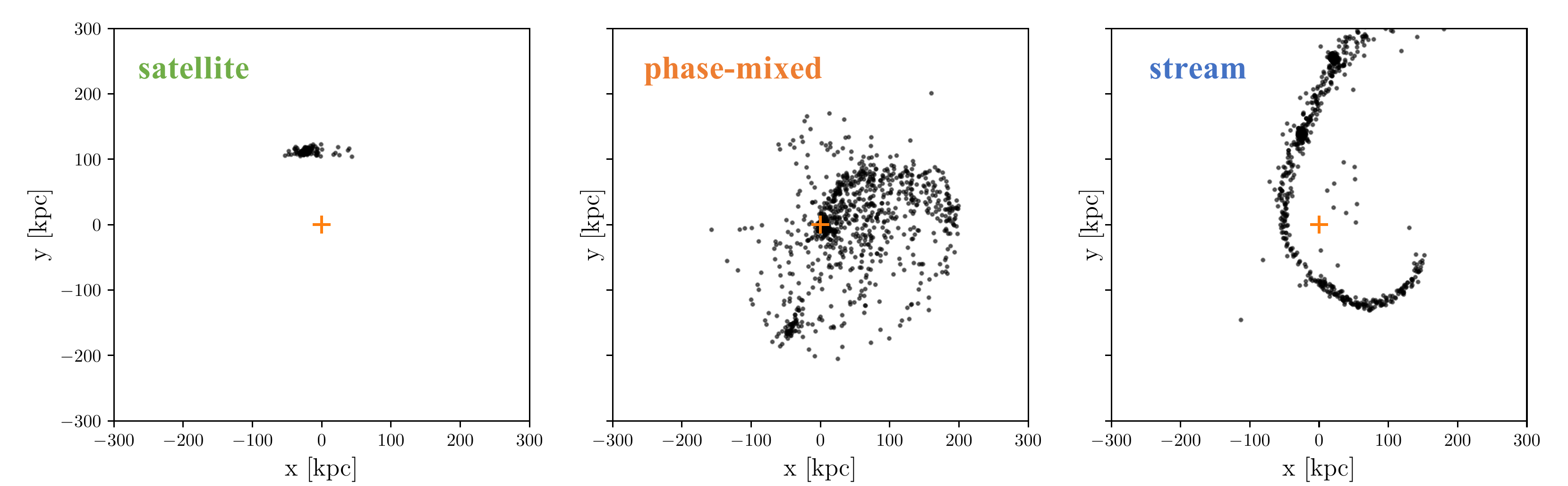}
\caption{Example objects, showing satellite galaxy (left, \texttt{m12i}), phase-mixed (middle, \texttt{m12b}) and stream (right, \texttt{m12b}) structures as determined by the criteria listed in \S\ref{sec:crit}. The satellite fails criterion (ii) (on the distance between star particles), while the phase-mixed object fails criterion (iii) (on the local velocity dispersion). Note the that distance spread of this particular satellite galaxy is $\sim 96$ kpc, which is less than the distance threshold of 120 kpc in criterion (ii). \label{fig:example_stream}}
\end{figure*}

We select stream-like objects in each cosmological simulation using a standardized set of criteria. This section explains the steps that were taken to identify these streams, and their progenitors, for further study.
\subsection{Stream Candidate Identification}
\label{subsec:tracking}
We select candidates with non-zero stellar mass that were self-bound objects at some point in the past (and may or may not be bound at present day). Since it takes several dynamical times for a bound satellite to turn into a stream, we select structures that are bound at any time 2.7--6.5 Gyr ago (corresponding to $z\sim 0.25-0.75$) and whose star particles are within the virial radius of the main galaxy at present day. We keep track of each selected object within this time range to eliminate duplicates. Extending the time window to earlier times, and increasing the tracking radius to larger than the present-day virial radius of the main galaxy, does not increase the total number of coherent streams. As we will show in \S \ref{subsec:disp-criterion-check}, most satellite galaxies the are phase-mixed at present day become unbound before $z\sim 0.75$, given that the underlying potential of the main galaxy in most simulations is still changing non-adiabatically up until this point \citep[see also][]{2020arXiv200103178S}. Moreover, at earlier times, there is not necessarily an unambiguous single host galaxy, so whether an object is accreted is not well-defined.

We track each substructure selected in this manner back in time to its formation, and forward to the point where it has merged with the main galaxy (or to the present day for objects that remain partially self-bound).  We discard massive subhalos that contain more star particles than our upper-limit threshold stated in \S\ref{sec:crit}. A subhalo is considered ``merged'' when \texttt{consistent-trees} no longer continuously tracks its center. Since the code allows for individual subhalos to be intermittently missed by the halo finder, assigning a provisional placeholder or ``phantom halo'' to keep track of its expected position, this effectively means that \texttt{Rockstar}, a 6-D phase-space finder, must be unable to find the subhalo in $>3$ consecutive snapshots. This is important for our study since halo finders often lose track of subhalos for a snapshot or two at pericenter; by bridging these gaps we can continue tracing the orbits of stream progenitors until a fairly advanced stage of tidal stripping. 

To collect all the star particles that ever belonged to a given building block, we first determine the time in which the bound satellite obtains maximum stellar mass. We collect all star particles that were assigned to the same subhalo by our preliminary tracking (see \S\ref{sec:sims}) in any snapshot in a 200-Myr window around this time. This allows us to recover nearly all star particles associated with the subhalo, including those that may only be marginally bound. These star particles are then tracked up until present day to determine their current positions. We carry out this procedure for all the stream candidates identified in each MW-mass host galaxy in the 7 isolated and 3 paired systems in our suite for a total of 13 hosts. This process governs $\sim15-75\%$ of the accreted stellar halo since the time when the mass ratio of the main halo to the second most massive halo is 3:1. This wide range corresponds to a diverse accretion history among our simulations \citep{2020MNRAS.497..747S}. 

\subsection{Stream Candidate Classification}\label{sec:crit}

We use the following three criteria to determine whether star particles assigned to each accreted satellite galaxy in the previous section form a stellar stream at present day:
\begin{enumerate}
    \item \textbf{Number of star particles.} The minimum number of star particles is greater than 120 and the maximum number of star particles is less than $10^5$.
    \item \textbf{Distance between star particles.} The maximum value of the pairwise separation between any two star particles in the group is greater than 120 kpc.
    \item \textbf{Local velocity dispersion.} The median of the local velocity dispersion of star particles, $\left\langle\sigma\right\rangle$, and the total stellar mass of the object,  $M_\star$, satisfies Equation \ref{eq:median_vel_dis}.
\end{enumerate}

We impose a lower bound on the number of star particles to ensure that all selected streams have enough star particles to be at least marginally resolved in the simulations. In the isolated simulations, since each evolved star particle has typical mass of $\sim 5000\Msun$, the streams and their progenitors have a lower mass limit of around $10^{6}\Msun$. This lower bound limits our study to simulated streams from dwarf galaxies at slightly above the mass of a classical dwarf satellite in the Milky Way. Stellar streams from globular clusters do not natively form in our simulations. An upper bound on the number of star particles is imposed to rule out objects that are even more massive than the Sagittarius stream, the most massive coherent stream known to exist in the real MW. The star particle mass resolution is slightly better in the paired simulation with each evolved star particle having a mass of $\sim 3000\Msun$. This essentially decreases the lower mass limit that we can probe by half, compared to the isolated simulations. The upper bound on the number of star particles limits the stellar masses of the objects to $\lesssim 10^{9} \Msun$, similar to the stellar mass of the most massive streams such as Sagittarius. This is to ensure that we limit to objects that are minor mergers.

Requiring a large maximum separation between star particles serves to eliminate dwarf galaxies that are still bound at present day. Bound satellite galaxies are compact in both position and velocity, hence, their maximum pairwise distances are small. The distance separation threshold of 120 kpc is comparable to the size of the main galaxy's halo, which is much larger than characteristic sizes of bound dwarf galaxies. This criterion selects streams that have wrapped at least partway around the host galaxy by the present day, and is especially robust at eliminating dwarf galaxies that are slightly tidally deformed but without prominent tails. If we were to relax the distance separation threshold to 90 kpc, we would have let in 9 more objects across all simulations. We have checked by eye that most of these additional objects are bound dwarf galaxies that are slightly tidally deformed. Only one object looks like a coherent stream with a relatively small orbital diameter of $\sim 100$ kpc, hence, evading our original separation threshold. 1--2 additional streams with small orbital radii that we might obtain from lowering the distance separation threshold would not change any conclusions in this manuscript.

The purpose of the local velocity dispersion cut is to eliminate candidate streams that are phase-mixed, which happens at a later stage as a stream is fully absorbed by the main galaxy. Specifically, we refer to structures for which the \emph{coarse-grained} phase-space distribution defined by the star particles associated with a particular stream is phase-mixed (i.e. approaches the overall phase-space distribution in the host halo). This criterion is thus resolution-dependent, as we discuss later in this section. Most known streams in the Milky Way and other galaxies are not phase-mixed, since phase-mixed objects are unlikely to be discovered and labeled outside of the local volume where 6 dimensional data (positions and velocities) are available. A few exceptions discovered in the Solar neighborhood using such data, such as the Helmi stream \citep{1999Natur.402...53H} and the Gaia Enceladus stream \citep{2018Natur.563...85H, 2018MNRAS.478..611B}, are thought to be phase-mixed. 

In our simulations, phase-mixed objects can be effectively ruled out by considering the median of the local velocity dispersion for all star particles along each candidate stream. The local velocity dispersion is computed using nearest neighbors in \emph{phase space} around each star particle, not neighbors in real space where star particles in different orbital phases might be selected if the stream has multiple wraps. When the nearest neighbor star particles in phase space are no longer neighbors in orbital phase, the velocity dispersion should increase sharply and we can consider the stream phase-mixed for our purposes. The phase space distance between the $i^{th}$ particle and the selected particle is defined to be the Mahlanobis distance
\begin{equation}
    d_i = \sqrt{\frac{(\vec{x}_i - \vec{x}_0)^2}{\sigma_x^2} + \frac{(\vec{v}_i - \vec{v}_0)^2}{\sigma_v^2}},
\end{equation}\\
where $\vec{x}_i$ is the position of the $i^{th}$ particle, $\vec{v}_i$ is the velocity of the $i^{th}$ particle, and $\sigma_x$ and $\sigma_v$ are the standard deviations of positions and velocities of all star particles in the stream. $d_i$ is thus a unitless quantity. The nearest neighbors to a given star, used to calculate its local velocity dispersion, are the stars with the 20 smallest $d_i$ for stream candidates with more than 300 star particles, and the 7 smallest $d_i$ otherwise. The local velocity dispersion is defined to be the velocity dispersion among these nearest neighbors.

Among the candidates that pass the number of star and the pairwise distance criterion, we identify phase-mixed candidates by eye in 4 of our 7 isolated simulations: \texttt{m12i}, \texttt{m12f}, \texttt{m12m} and \texttt{m12c} to calibrate this criterion. We compute the local velocity dispersion for all star particles in each candidate stream, estimating the values at $16^{th}$, $50^{th}$ and $83^{rd}$ percentile. Figure \ref{fig:dispersion_cut} shows these values for each candidate in all 4 simulations with respect to their total stellar mass. The marker represents the median value. The error bars represent the $16^{th}$ and $83^{rd}$ percentile. Phase-mixed candidates are shown in orange, while non-phase-mixed candidates are shown in blue. The phase-mixed candidates have systematically higher median local velocity dispersions compared to the non-phase-mixed ones. We use a linear kernel Support Vector Machine (SVM), implemented in the \texttt{scikit-learn} package, to determine the best line that separates the two groups \citep{Cortes:1995:SN:218919.218929, scikit-learn}. The algorithm finds that the hyperplane that maximizes the width of the gap between the two groups is
\begin{equation}\label{eq:median_vel_dis}
    \left\langle\sigma\right\rangle = -5.28\log \left(\frac{ M_\star}{M_\odot}\right) + 53.55,
\end{equation}
where $\left\langle\sigma\right\rangle$ is the median of the local velocity dispersion and $M_\star$ is the total stellar mass. Candidates that lie above this line are considered phase-mixed. The local velocity dispersion cut is stellar mass dependent since lower mass candidates have fewer star particles which result in higher estimated local velocity dispersion. This is shown by the negative slope of the cut, and is a reflection of the resolution dependence of this criterion. However, the resolution dependence is quite weak, as shown by the very shallow slope of the cutoff with stellar mass. In particular, it can still be applied to the paired simulations we analyze, which have mass resolution about twice that of the isolated simulations (3500 versus 7100 $M_\odot$). We confirm by eye that the stream candidates in the paired simulations that lie above the threshold are phase-mixed or borderline phase-mixing.

Table \ref{table:summary} shows the number of streams that pass all three criteria for phase-coherent tidal streams in each simulation, along with the main halo's total mass, stellar mass, and virial radius. There are more streams in the paired simulations because of their better particle resolution. The number of streams in both sets of simulations are comparable if we only consider streams in the paired simulations that are more massive than the lower mass limit of the unpaired simulations at $\sim 10^6\Msun$. Example of objects that are classified by our criteria as satellite, stream, and phase-mixed are shown in Figure \ref{fig:example_stream}. To summarize our classification process: all objects that we consider pass criterion (i) (on the number of star particles). Satellites fail criterion (ii) (on the distance between star particles). Phase-mixed objects pass criterion (ii), but fail criterion (iii) (on the local velocity dispersion). Streams pass all the criteria. These important definitions are also summarized in Table \ref{tab:defn}.

\subsection{Selecting Recent/First Infall Satellites}
Since we only search for self-bound objects within the host halo's virial radius between 2.7--6.5 Gyr ago when we look for present-day coherent streams, our sample of satellite dwarf galaxies are incomplete, potentially missing some satellites that fell in more recently. To include these recent infall satellites, we first identify self-bound objects with star particles within the each host's virial radius at z=0. These luminous bound objects are then compared to our set of previously identified satellites and streams (with still visible bound part at $z=0$). All phase-mixed objects no longer have identifiable bound parts. We select out extra, non-overlapping objects as recent infall satellites. We follow steps in \S\ref{subsec:tracking} to collect most star particles that once belong to these additional satellites. 34 additional recent infall satellites are identified across all of the simulations.

\begin{deluxetable}{lccl}
\tablehead{
\colhead{sim name} & \colhead{$m_\mathrm{200m}$[M$_\odot$]} & \colhead{$r_\mathrm{200m}$[kpc]}& \colhead{N}}
\startdata
m12i & $1.18\times10^{12}$ & 336 & 9\\
m12f & $1.71\times10^{12}$ & 380 & 8\\
m12m & $1.58\times10^{12}$ & 371 & 8 \\
m12c & $1.35\times10^{12}$ & 351 & 7 \\
m12b & $1.43\times10^{12}$ & 358 & 8 \\
m12r & $1.10\times10^{12}$ & 321 & 3 \\
m12w & $1.08\times10^{12}$ & 319 & 3 \\
Romeo &$1.32\times10^{12}$& 341 &13\,[10]\\
Juliet &$1.10\times10^{12}$& 321&12\,[6]\\
Romulus &$2.08\times10^{12}$& 406&9\,[6]\\
Remus &$1.22\times10^{12}$& 339&8\,[5]\\
Thelma &$1.43\times10^{12}$ & 358& 10\,[9]\\
Louise &$1.15\times10^{12}$& 333& 8\,[8]\\
\enddata
\caption{Total mass ($m_\mathrm{200m}$, DM+star+gas), virial radius ($r_\mathrm{200m}$) and number of stellar streams that pass all the criteria for phase-coherent tidal stream for each simulated Milky Way-like galaxies. The numbers in the square brackets are the effective numbers of stellar streams in the paired galaxy simulations that have mass greater than the lower bound mass we set for the unpaired simulations. Here, the total mass refers to the total mass enclosed within $r_\mathrm{200m}$, a radius containing 200 times the mean background matter density. \label{table:summary}}
\end{deluxetable}

\begin{deluxetable*}{cl}

\tablecaption{Important Definitions}
\tablehead{& \textit{Stream Candidate Identification} (refer to \S\ref{subsec:tracking})} 

\startdata
Stream candidate & An object that is bound between 2.7--6.5 Gyr ago and is within present day's \\
& virial radius of the host.\\
  \hline \hline
  & \textit{Candidate Classification} (refer to \S\ref{sec:crit}) \\
  \hline
Satellite & A candidate that has between 120--$10^5$ star particles at present day, but has maximum \\
& pairwise distances between star particles less than 120 kpc. \\
Phase-mixed object &  A candidate that has between 120--$10^5$ star particles at present day, has maximum \\
& pairwise distances between star particles greater than 120 kpc, and has median local \\
& velocity dispersion greater than Equation \ref{eq:median_vel_dis}. \\
Stream & A candidate that has between 120--$10^5$ star particles at present day, has maximum \\
& pairwise distances between star particles greater than 120 kpc, and has median local \\
& velocity dispersion less than Equation \ref{eq:median_vel_dis}. \\
  \hline \hline
  & \textit{Timescales} \\
  \hline
  $\tau_\textrm{infall}^{i}$ & First infall look-back time, measured at the \emph{first} virial radius crossing.\\
  $\tau_\textrm{infall}^{f}$ & Last infall look-back time, measured at the \emph{last} virial radius crossing.\\
  $\tau_\textrm{stream}$ & Stream-formation look-back time, measured when the stellar part of the progenitor starts to \\ 
  & become unbound (refer to \S\ref{subsec:infall-form}). This is also the age of each stream/phase-mixed object.\\
  $\tau_\textrm{peak}$ & Peak stellar mass look-back time, measured when the bound part of the progenitor contains \\
  & peak stellar mass throughout its evolution.
\enddata
\tablecomments{For objects that only cross the virial radius once, $\tau_\textrm{infall}^{i} = \tau_\textrm{infall}^{f}$.\label{tab:defn}}
\end{deluxetable*}

\section{Properties of stellar stream progenitors} \label{sec:prop}
Here we summarize the properties and statistics of the progenitors of all the streams that were selected as described in Section \ref{sec:select}. These dwarf-galaxy progenitors of present-day ($z=0$) tidal streams are characterized at the time when they have maximum stellar mass, unless otherwise stated.

\subsection{Mass Function}
\label{subsec:massfunc}
\begin{figure}
\plotone{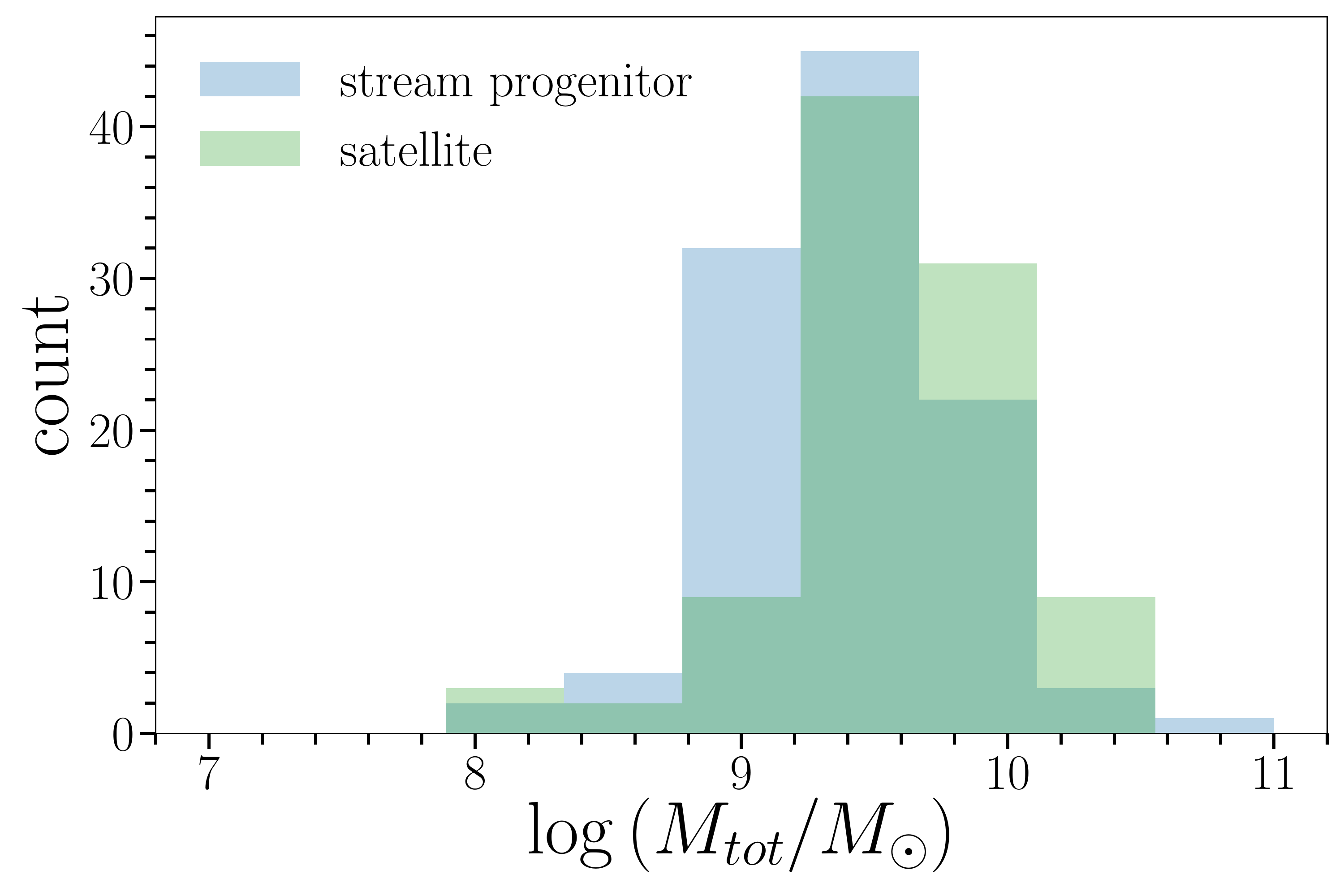}
\caption{Total mass functions for satellite dwarf galaxies (green) and for stellar stream progenitors (blue), cumulative across all simulations. The total mass is evaluated at the peak stellar mass timescale of each satellite and progenitor. \label{fig:stat_mass}}
\end{figure}

\begin{figure}
\plotone{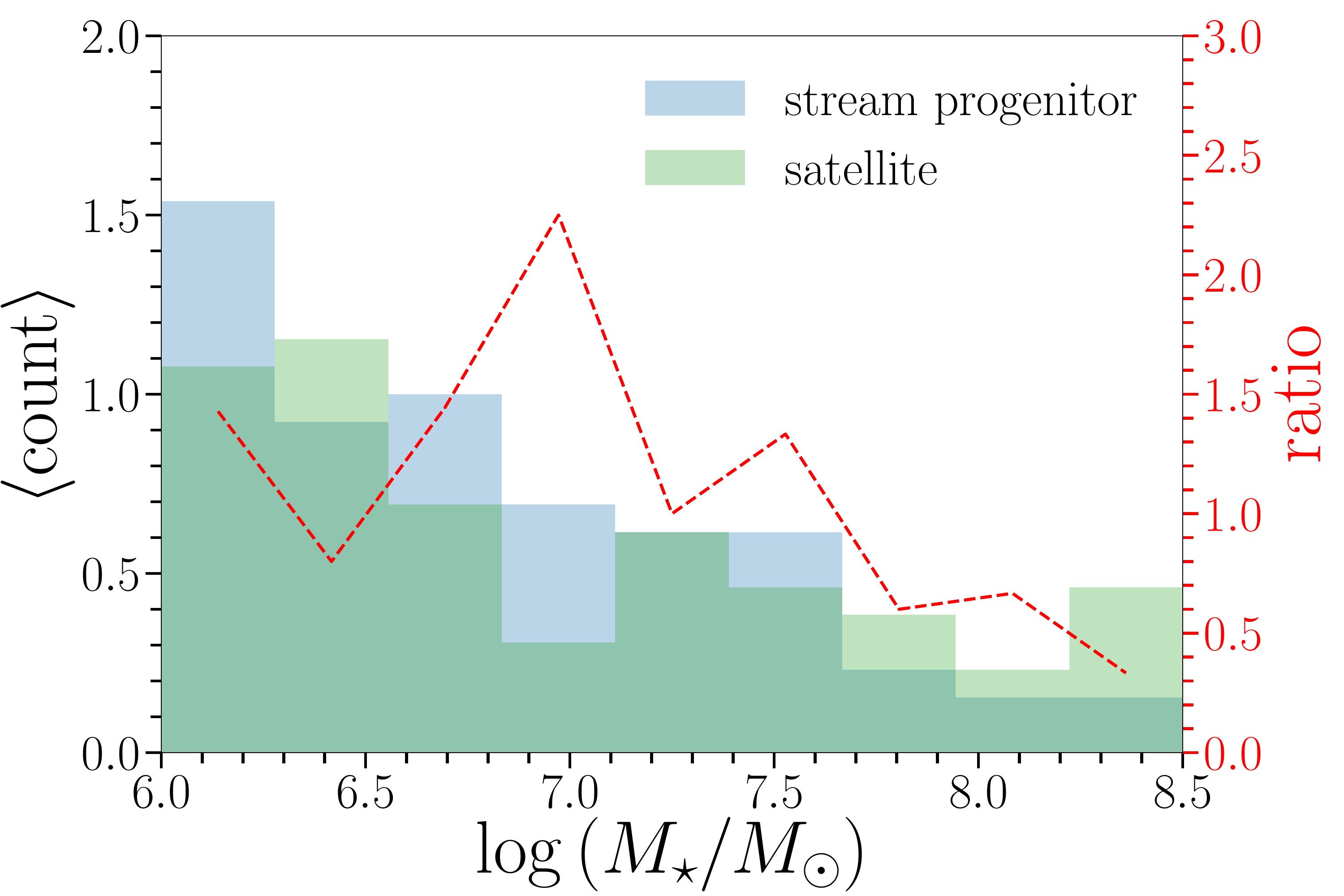}
\caption{Stellar mass functions for satellite dwarf galaxies (green) and for stellar stream progenitors (blue). These are cumulative across all simulations, normalized by the total number of hosts; hence, $\left< \mathrm{count} \right>$ represents the average number of objects per host in any given stellar mass bin. The stellar mass is evaluated at the peak stellar mass timescale of each satellite and progenitor. The ratio of the progenitor to satellite galaxy is shown by the dashed line (red). \label{fig:stat_stellar}}

\end{figure}
The total and stellar mass functions of stream progenitors and satellites are alike, which show that there is essentially no preferred range in total mass or stellar mass for which satellite galaxies are more likely to turn into stellar streams.
Figure \ref{fig:stat_mass} compares the total mass (stars, gas, and DM) distribution of the stream progenitors (blue) with the total mass distribution for satellite galaxies (green), both of which are evaluated at the peak stellar mass timescale of each object. These distributions are cumulative across all simulations. Both distributions show similar features. The majority of the stream progenitors and satellites have total mass of $\sim 10^9 - 10^{10} \Msun$. However, note that these total mass distributions are localized and do not span the entire mass spectrum. This is primarily due to the lower bound and upper bound we set on the number of star particles in the object that we track.

Figure \ref{fig:stat_stellar} is similar to Figure \ref{fig:stat_mass}, but shows the distribution of stellar mass, evaluated at the peak stellar mass timescale of each object, instead of total mass and is only plotted over the span of stellar mass range that we have completed sample which corresponds to our criterion (ii) for the number of star particles. The stellar mass function of the stream progenitors is shown in blue, while the stellar mass function of satellites is shown in green. Both distributions have the most objects at low stellar mass, and the number decreases at higher stellar masses. The ratio of the number of stream progenitor to satellite at each stellar mass bin is shown by the dashed line (red). At stellar mass $\sim 10^{7.5}\Msun$, there are comparable number of stream progenitors and satellites. There are more satellite galaxies at higher stellar mass (ratio $< 1$), while there are more stream progenitors at lower stellar mass (ratio $> 1$). The decrease in the relative numbers of stream progenitors to satellites at high stellar mass can be attributed to the shorter dynamical lifetimes for massive streams due to experiencing higher dynamical friction.

\subsection{Infall and Stream-formation Timescales}
\label{subsec:infall-form}
The infall and formation times of stellar streams are crucial in modeling and understanding the origin of real observed streams. In this work, we define the \emph{infall time}, $\tau_{\mathrm{infall}}$, as the time before present day that each progenitor crosses the virial radius of the main galaxy. The virial radius of the main galaxy is determined independently for each snapshot in each simulation, and increases monotonically with time. We find that about half of stream progenitors cross the virial radius multiple times before complete disruption. In these cases, we distinguish between $\tau_{\mathrm{infall}}^{i}$, first infall, and $\tau_{\mathrm{infall}}^f$, final infall.  For systems with a single virial radius crossing, $\tau_{\mathrm{infall}}=\tau_{\mathrm{infall}}^i=\tau_{\mathrm{infall}}^f$.

The \emph{stream-formation time}, $\tau_{\mathrm{stream}}$, refers to the time when the stellar part of the dwarf galaxy progenitor starts to become unbound, and star particles that once belonged to the progenitor start to stretch out along the orbit. In this paper, we use the terms ``stream-formation time'' and ``formation time'' interchangeably. We determine $\tau_{\mathrm{stream}}$ by computing the moment-of-inertia tensor for the progenitor, $I_{ij}$:
\begin{equation}
    I_{ij} = -\sum_{n=1}^N m_n x_{i,n} x_{j,n},
\end{equation}
for $i \neq j$ and,
\begin{equation}
    I_{11} = \sum_{n=1}^N m_n \left(x_{2,n}^2 +  x_{3,n}^2\right),
\end{equation}
\begin{equation}
    I_{22} = \sum_{n=1}^N m_n \left(x_{1,n}^2 +  x_{3,n}^2\right),
\end{equation}
\begin{equation}
    I_{33} = \sum_{n=1}^N m_n \left(x_{1,n}^2 +  x_{2,n}^2\right),
\end{equation}
for $i=j$, where $i, j = 1,2,3$ and $N$ is the total number of star particles in each progenitor. $x_{i,n}$ is the position of the $n^{th}$ star particle along the $i^{th}$ direction, and $m_n$ is the mass of the $n^{th}$ star particle. 

\begin{figure*}
\plotone{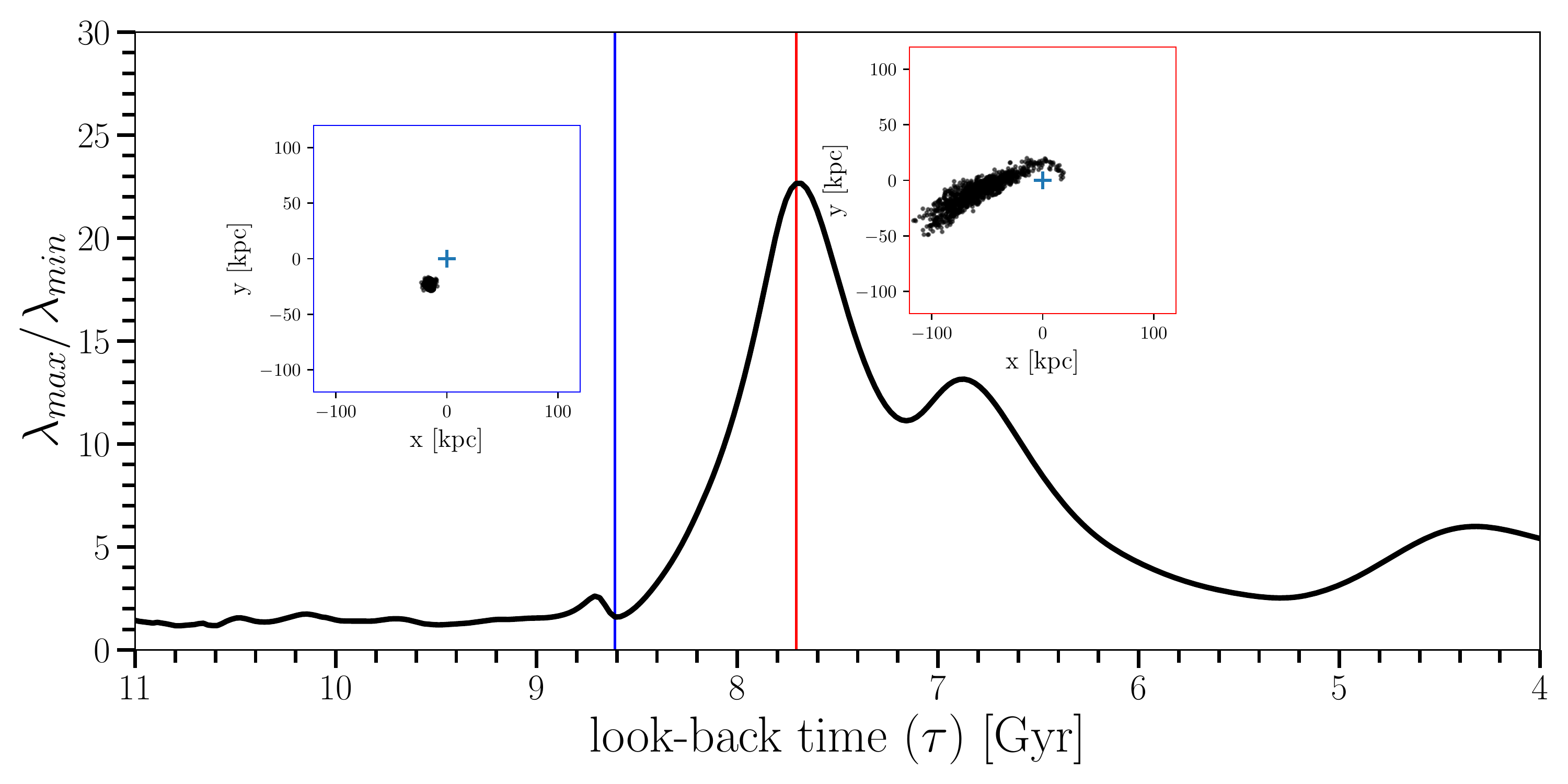}
\caption{A typical evolution of the ratio of the maximum to minimum eigenvalue ($\lambda_{max}/\lambda_{min}$) of a stream progenitor's moment-of-inertia tensor as a function of look-back time. The graph shown is for an example stream progenitor from \texttt{m12i}. The red vertical line marks the global peak of the entire evolution. The blue vertical line marks the first local minimum just before the global peak and is the timescale that characterizes the stream-formation look-back time ($\tau_{\mathrm{stream}}$) for this example stream progenitor. The two panels show 2-D projections of the progenitor at these two timescales. \label{fig:st_form}}

\end{figure*}

We compute the three eigenvalues $\lambda$ of $I_{ij}$ and determine the maximum ($\lambda_{max}$) and minimum ($\lambda_{min}$) values. The ratio $\lambda_{max}/\lambda_{min}$ generally starts near 1 (while the object is self-bound), dips slightly during tidal compression, increases sharply as tidal disruption begins and eventually approaches 1 again as the stream wraps around the galaxy. A typical evolution of the eigenvalue ratio for an example stream progenitor is shown in Figure \ref{fig:st_form}. The local minimum just before the maximum of $\lambda_{max}/\lambda_{min}$ is defined as $\tau_{\mathrm{stream}}$. We limit the search for  $\tau_{\mathrm{stream}}$ to times after the first crossing of the virial radius of the main galaxy, in order to ensure that the elongation is due to interaction with the main galaxy and not early assembly of the progenitor. 

\begin{figure}
\plotone{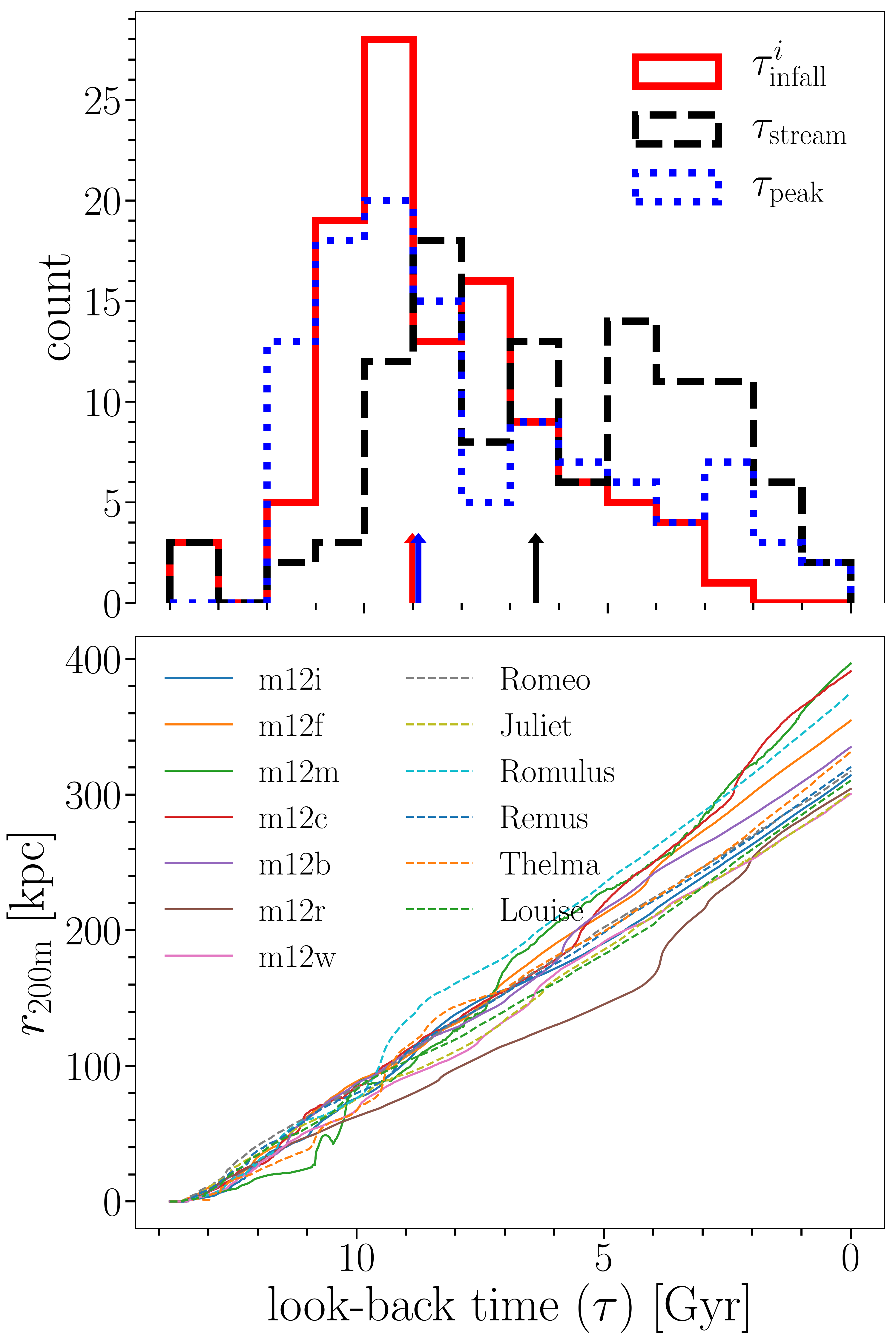}
\caption{Top: first infall look-back time $\tau_{\mathrm{infall}}^i$ (solid red), stream-formation look-back time $\tau_{\mathrm{stream}}$ (dashed black) and peak stellar mass look-back time $\tau_{\mathrm{peak}}$ (dotted blue) of stellar stream progenitors across all 13 simulations. These are shown in look-back time $\tau$. The three distributions are not entirely identical, as these timescales do not happen at fixed ordering as discussed in section \S\ref{subsec:ordering}. Bottom: time evolution of the virial radii $r_\mathrm{200m}$ of the host galaxies in isolated simulations (solid lines) and paired simulations (dashed lines). \label{fig:stat_infall}}

\end{figure}

The top panel of Figure \ref{fig:stat_infall} shows histograms of the first infall look-back time $\tau_{\mathrm{infall}}^i$ (solid red), stream-formation look-back time $\tau_{\mathrm{stream}}$ (dashed black) and peak stellar mass look-back time (dotted blue) of stellar stream progenitors, while the bottom panel shows the virial radius of each main galaxy over the same timescale. Many stream progenitors fall into the main galaxy relatively early on: over 6 Gyr ago, when the virial radii of the main galaxies are approximately half of their present day values. However, most streams do not form as soon as their progenitors fall into the main galaxy, as shown by the significant shift between the two distributions. In some cases, it can take several Gyr before the progenitor is tidally disrupted, depending on its orbit. The stellar streams that have formed most recently are all either in very radial orbits or still have visible self-bound parts. Lastly, the distribution for the peak stellar mass timescale is also slightly shifted from both the infall and stream-formation timescales.


Our method of estimating the stream-formation time for each progenitor is purely geometrical; it does not use any dynamical information for each star particle. This strategy is intended to mimic observational identification of streams in large surveys, which generally uses only positional information. However, one drawback of our method is that it cannot reliably determine $\tau_{\mathrm{stream}}$ for members of an infalling group whose constituent objects are not fully merged after the first infall. Stream-formation times for such objects, which can tidally disrupt each other prior to infall into the host galaxy, are somewhat ambiguous. Out of the total of 106 streams, there are 13 objects that we mask as group infalls (roughly one per host). Table \ref{tab:defn} contains definitions of these important timescales.

\subsection{Ordering of infall, quenching, and stream formation timescales}
\label{subsec:ordering}

\begin{figure*}
\plotone{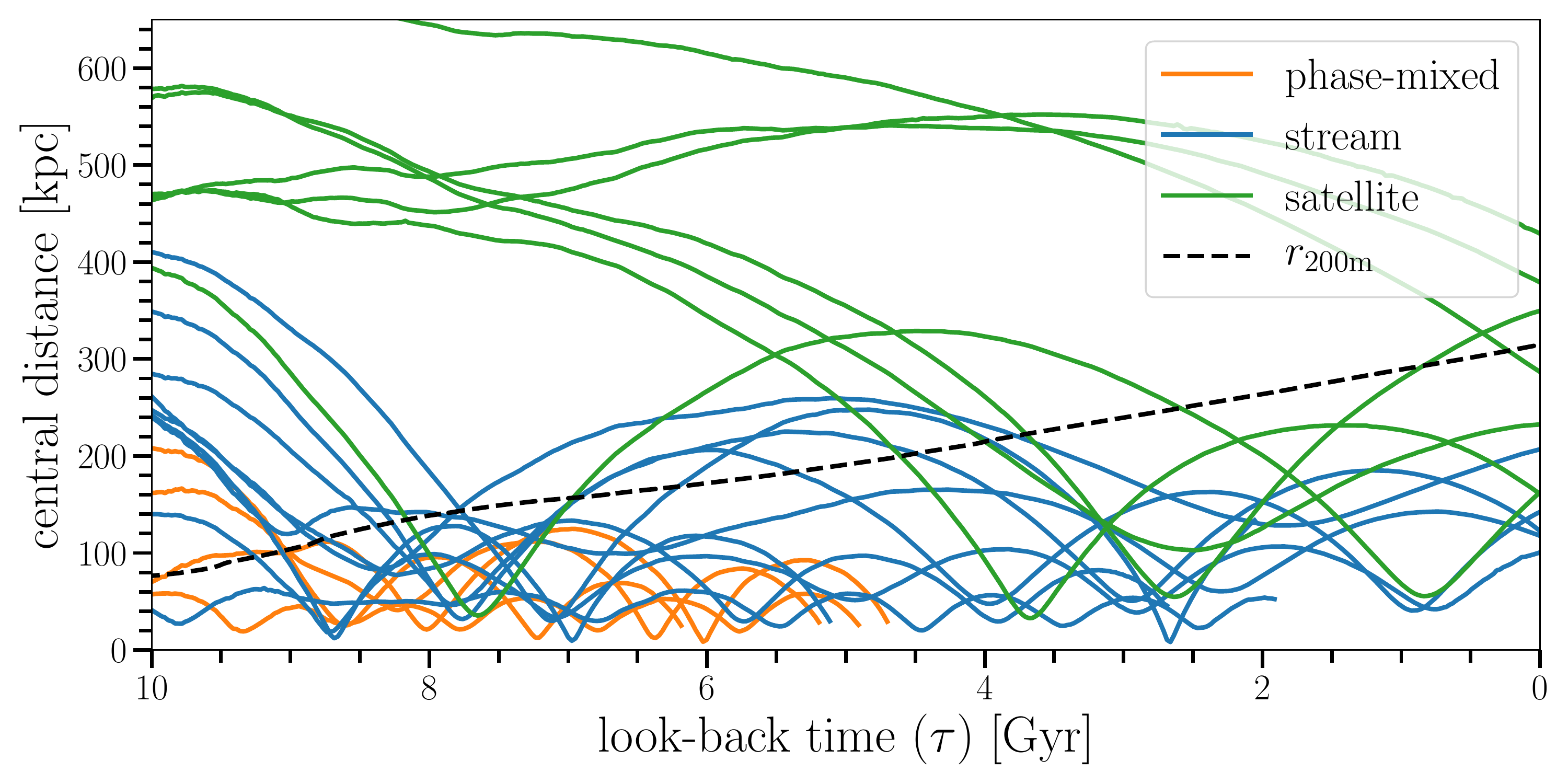}
\caption{Assembly history of the stellar halo of \texttt{m12i} plotting object's central distance as a function of a look back time $\tau$. Line colors represent object classifications at $z=0$ based on criteria in \S\ref{sec:crit}. Each line terminates when the object no longer has an identifiable self-bound part, or is drawn up to $z=0$ if the object is still partially self-bound at present day. The virial radius ($r_\mathrm{200m}$) of the main halo is shown as a black dashed line. \label{fig:assembly}}

\end{figure*}

Figure \ref{fig:assembly} shows the assembly history of present day dwarf galaxies, stellar streams and phase-mixed objects, identified by the selection criteria in \S\ref{sec:crit}, into the stellar halo of the isolated simulation \texttt{m12i}. This history is typical of our simulated systems without a relatively recent major merger. Phase-mixed components (orange lines) exclusively fall in very early and are all completely disrupted by 5 Gyr ago, while present-day dwarf satellite galaxies all have their last crossing of the virial radius after this time. The streams, meanwhile, probe basically the entire lifetime of the galaxy. Many of the more recently disrupted streams, as well as several of the satellite galaxies, have experienced multiple virial radius crossings, suggesting that their orbits could potentially be influenced by the local environment \emph{beyond} the virial radius, especially in the paired environment. The transfer of an object from one host to the other does happen in the paired simulations, although they are not common ($\sim$1--2 objects per simulation). The dynamical properties of stream progenitors in the paired simulations are largely only confined within each host, similar to the isolated simulations.

 The standard picture of hierarchical assembly in galactic stellar halos includes an implicit ordering of several important timescales in the life of each progenitor of a stellar stream. In this picture, each satellite galaxy crosses the virial radius of the host at $\tau_{\mathrm{infall}}$, has its star formation quenched (if it was not already) by the environment in the halo at $\tau_{\mathrm{peak}}$, and then is tidally disrupted to form a stream at $\tau_{\mathrm{stream}}$. From this picture, we would generally expect $\tau_{\mathrm{infall}} > \tau_{\mathrm{peak}} > \tau_{\mathrm{stream}}$ for more massive satellites with sustained star formation, and $\tau_{\mathrm{peak}} \geq \tau_{\mathrm{infall}} > \tau_{\mathrm{stream}}$ for less massive satellites quenched prior to infall by reionization or stellar feedback, with all ages measured back in time from the present day. Following this reasoning, the measured ages of stars in streams (which are $\geq \tau_{\mathrm{peak}}$) are sometimes used as an upper bound on the stream age $\tau_{\mathrm{stream}}$. Figure \ref{fig:stat_infall} shows that statistically speaking, $\tau_{\mathrm{peak}} \sim \tau_\mathrm{infall}^i\geq \tau_{\mathrm{infall}}^f \gtrsim \tau_{\mathrm{stream}}$ when the entire sample is considered. However, the spread of all three timescales is extremely broad, and the size of the shift between $\tau_{\mathrm{peak}}$ and $\tau_{\mathrm{infall}}$ nearly vanishes if we consider $\tau_{\mathrm{infall}}^i$ rather than $\tau_{\mathrm{infall}}^f$.

\begin{figure*}
\plotone{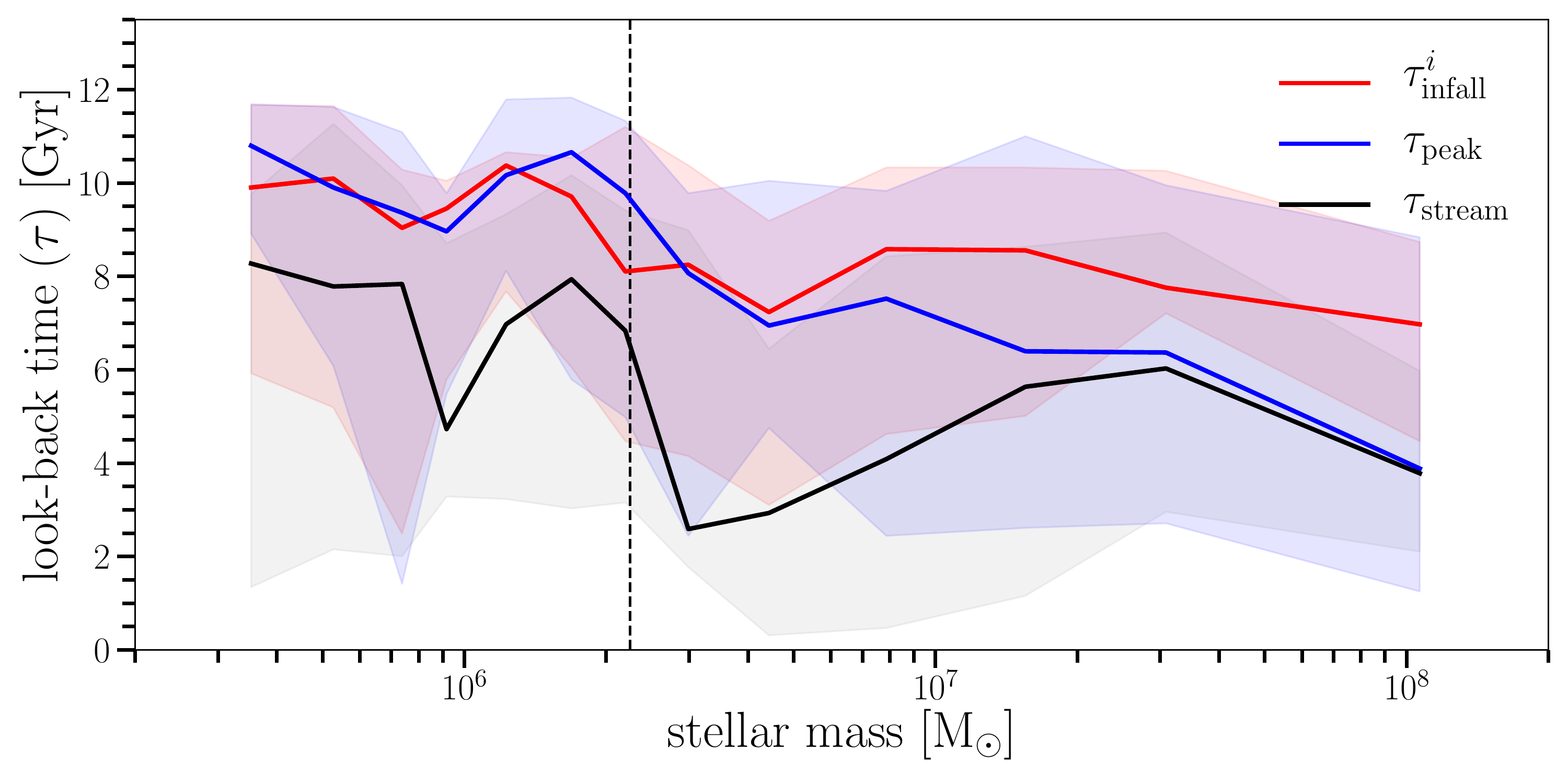}
\caption{Ordering of different timescales in the life of stream progenitors. Each timescale is binned in increasing stellar mass bins such that there are 7 progenitors in each bin. The lines represent the median value and shaded region encapsulate minimum and maximum values witin a given bin. The red represents the first infall look-back time $\tau_\mathrm{infall}^i$. The blue represents the peak stellar mass time, $\tau_\mathrm{peak}$. The black represents the stream-formation look-back time, $\tau_\mathrm{stream}$, described in \S\ref{subsec:infall-form}. The black dashed vertical line is the median stellar mass of all progenitors ($\sim 2.25\times 10^{6}\Msun$). \label{fig:timing}}

\end{figure*}

\begin{figure*}
\plotone{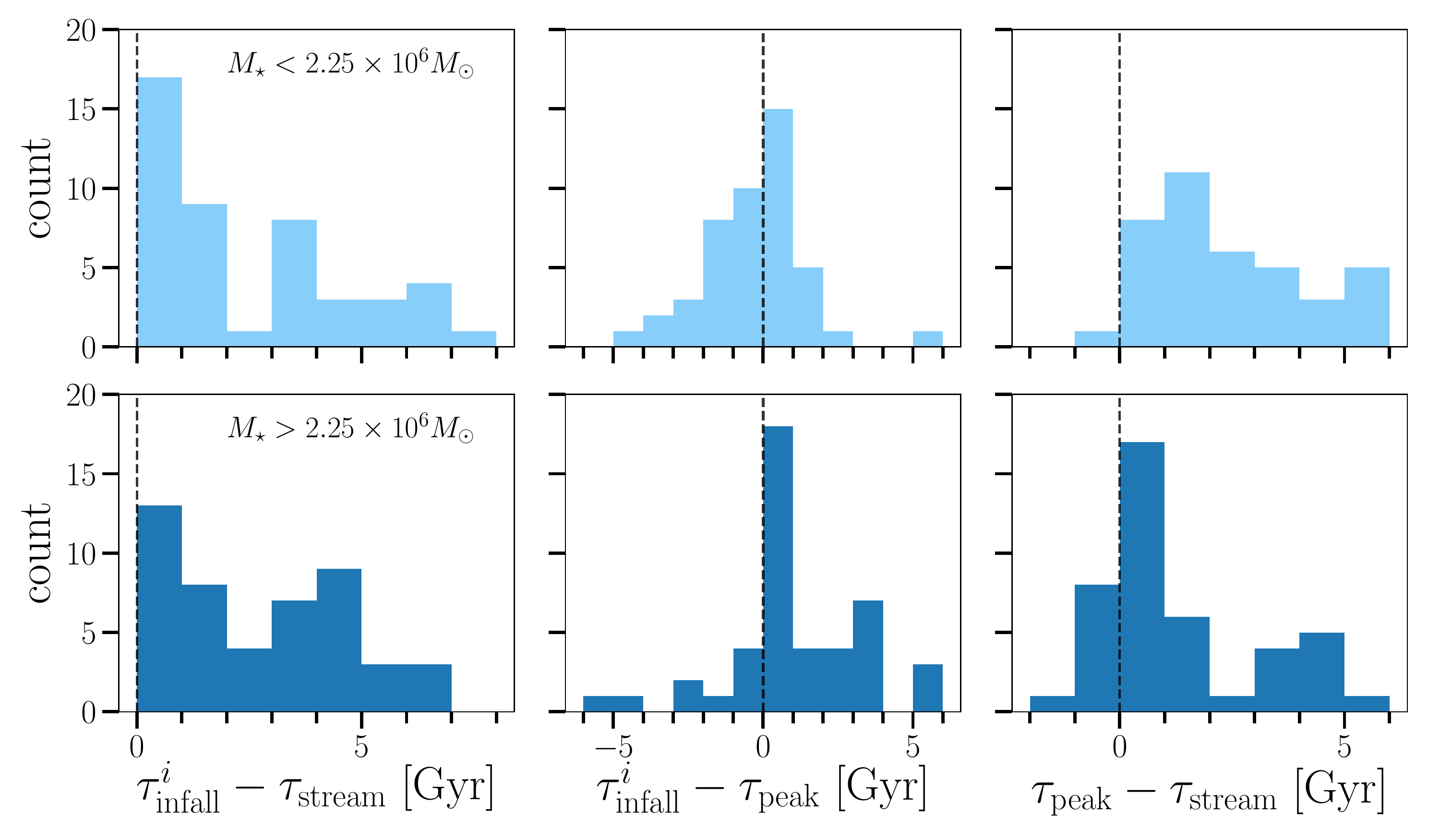}
    \caption{Relative values of different look-back timescales in the life of low-mass (top row; light blue) and high-mass (bottom row; dark blue) stream progenitors. The columns represent different pairs of timescales. Left is the infall look-back time relative to the stream-formation look-back time ($\tau_\mathrm{infall}^{i} - \tau_\mathrm{stream}$). Middle is the infall look-back time relative to the peak stellar mass look-back time ($\tau_\mathrm{infall}^{i} - \tau_\mathrm{peak}$). Right is the peak stellar mass look-back time relative to the stream-formation look-back time ($\tau_\mathrm{peak} - \tau_\mathrm{stream}$). The positive values translate to the first timescale occurring before the second timescale. The vertical dashed lines divide the positive and negative relative values.}
    \label{fig:ordering-pair}
\end{figure*}

\begin{figure*}
\plotone{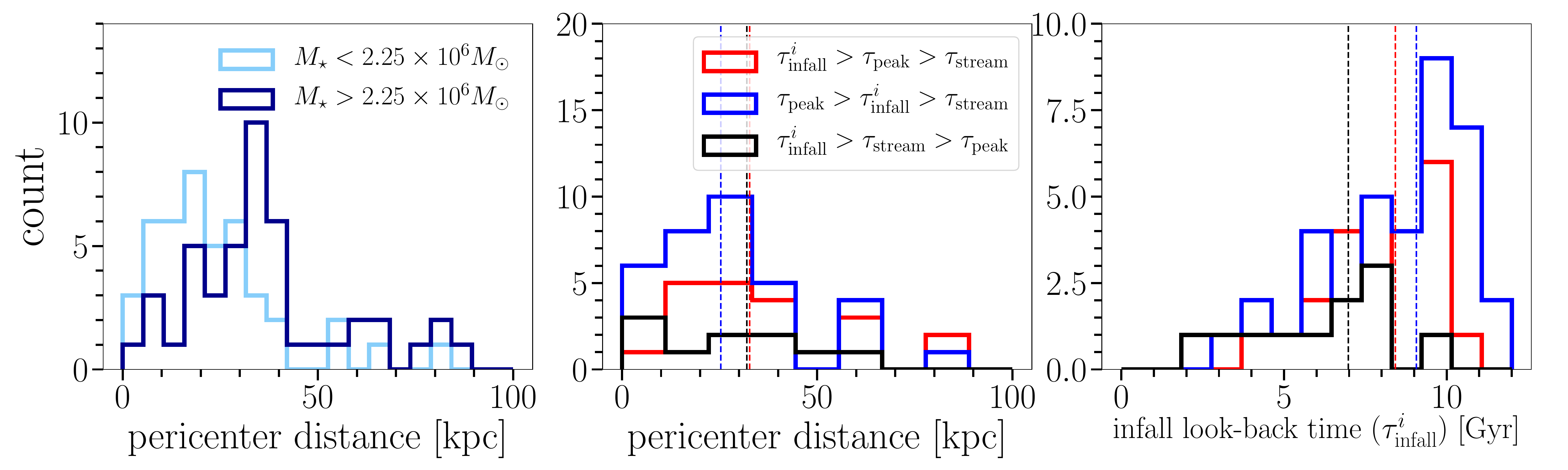}
\caption{Left: Distribution of the pericenter distances for the low-mass progenitors (light blue) and high-mass progenitors (dark blue). Present day high-mass coherent streams are more likely to have larger pericentric distances compared to low-mass coherent streams. Middle: Distribution of the pericenter distances for progenitors (low-mass and high-mass combined) with three different ordering of timescales that they experience in their lifetime (see \S\ref{subsec:ordering}). The dashed vertical lines are the median of the distributions. Right: Distribution of the infall time ($\tau_\mathrm{infall}^i$) for progenitors with three different ordering of timescales. Note that there are 5 progenitors with pericenter distance $>100$ kpc, which are not shown in the left and middle figures.}
\label{fig:ordering-pericenter}
\end{figure*}

On closer examination, we find that this picture is not universally applicable to the progenitors of streams in our simulations. Figure \ref{fig:timing} shows the relative times of these different events for all stream progenitors in our sample, arranged by stellar mass. We bin the sample in increasing stellar mass bins such that each bin contains 7 progenitors. The median value in each bin are represented by lines and the shaded regions encapsulate minimum and maximum spreads. The infall time, $\tau_{\mathrm{infall}}$, is shown in red. The stream-formation time $\tau_{\mathrm{stream}}$ and peak stellar mass time $\tau_{\mathrm{peak}}$ are shown as black and blue, respectively. The median stellar mass of the entire sample $\sim 2.25\times 10^{6}\Msun$ is shown by the dashed vertical line. The median stellar mass is used to divide the sample into two subgroups: the high-mass and low-mass progenitors. This stellar mass cut is supported by the star formation histories of the Local Group dwarf galaxies: the dwarf satellite Sculptor is the most massive satellite known for which models suggest early quenching of star formation \citep{2014ApJ...789..148W}. Sculptor is estimated to have $M_\star \sim 2.3\times 10^6 \Msun$ \citep{2012AJ....144....4M}, consistent with our separation between high- and low-mass progenitors. Additionally, Figure \ref{fig:ordering-pair} shows relative values of all the pairs between the three timescales. The three distributions are plotted: $\tau_\mathrm{infall}^{i} - \tau_\mathrm{stream}$, $\tau_\mathrm{infall}^{i} - \tau_\mathrm{peak}$ and $\tau_\mathrm{peak} - \tau_\mathrm{stream}$. The top row is the low-mass group and the bottom row is the high-mass group. Note that the bi-modalities in the distributions of $\tau_\mathrm{infall}^{i} - \tau_\mathrm{stream}$ is real and can be attributed to the fact that if the progenitor does not form a stream during the first pericenter, it is unlikely to form a stream until subsequent pericenters.

For these two groups in mass, we examine three different orderings of significant events:
\begin{enumerate}
    \item $\tau_{\mathrm{infall}}^i > \tau_{\mathrm{peak}} > \tau_{\mathrm{stream}}$. The progenitor first falls into the main halo, then has its star formation quenched, then forms a stream. This is the standard picture for high-mass progenitors.
    \item $ \tau_{\mathrm{peak}} > \tau_{\mathrm{infall}}^i > \tau_{\mathrm{stream}}$. The progenitor first stops forming stars, then falls into the main halo, then forms a stream. This is the standard picture for low-mass progenitors.
    \item $\tau_{\mathrm{infall}}^i > \tau_{\mathrm{stream}} > \tau_{\mathrm{peak}}$. The progenitor first falls into the main halo, then begins to form a stream while still forming stars \emph{at a higher rate than they are stripped by tides}, then reaches maximum stellar mass before being completely tidally disrupted.
\end{enumerate}
In this subsection, we use $\tau_{\mathrm{infall}}^i$ as the infall time, since this marks the time when the progenitor first enters the main halo environment, which is presumed to quench its star formation.

For the high-mass group, out of 47 objects, 26 are in the first scenario, which is expected for this group, ($\sim 55\%$), 9 are in the second scenario ($\sim 19\%$) and 10 are in the third scenario ($\sim 21\%$). Two objects have equal $\tau_{\mathrm{stream}}$ and $\tau_{\mathrm{peak}}$, hence, are excluded. For the low-mass group, out of 46 objects, 20 are in the first scenario ($\sim 43\%$), 24 are in the second scenario, which is expected for this group, ($\sim 52\%$) and 2 are in the third scenario ($\sim 4\%$).

Our intuition for the ordering of timescales for high-mass and low-mass progenitors is thus approximately correct: around half of the members of each group follow our expected picture. However, there are large discrepancies within each group. A few progenitors in the high-mass group have their star formation quenched before their first infall. In the low-mass group, almost half of the progenitors still form star particles after their first infall. A significant fraction of the high-mass progenitors still have a high star formation rate even after undergoing tidal stripping, while this situation very rarely occurs for low-mass progenitors. In future work, Samuel et al. in prep. will examine the timescales and physical processes involved in satellite quenching in detail.

In Figure \ref{fig:ordering-pericenter}, we study the correlation between the pericenter distance of each progenitor and the timescale ordering that it experiences. Dividing the low-mass (light blue) and high-mass (dark blue) groups, the left plot shows the distribution of the pericenter distances. To remain coherent at present day, progenitors in the low-mass group overall have smaller pericenter distance with the peak at $\sim 15-20$ kpc, while the high-mass group peaks at $\sim 30-35$ kpc. This suggests that high-mass progenitors become phase-mixed with the host galaxy faster. Thus, high-mass progenitors with low pericenter distance are less likely to remain coherent at $z=0$. The middle and right panels show the distribution of the pericenter distance and infall time of progenitors in three scenarios of the timescales ordering, combining both low-mass and high-mass groups. The first ordering scenario ($\tau_{\mathrm{infall}}^i > \tau_{\mathrm{peak}} > \tau_{\mathrm{stream}}$) is shown in red. The second ordering scenario ($ \tau_{\mathrm{peak}} > \tau_{\mathrm{infall}}^i > \tau_{\mathrm{stream}}$) is shown in blue. The third ordering scenario ($\tau_{\mathrm{infall}}^i > \tau_{\mathrm{stream}} > \tau_{\mathrm{peak}}$) is shown in black. The median value of each distribution is illustrated by a dashed vertical line. The pericenter distributions are not clearly separable, with all three having similar median pericenter distance. However, they appear more distinct in the $\tau_\mathrm{infall}^i$ projection. Progenitors in the second ordering fall in the earliest, followed by progenitors in the first ordering and the third ordering, respectively. This trend can be explained by the connection between $\tau_\mathrm{inafll}^i$ and stellar mass of the progenitor. A small negative trend in Figure \ref{fig:timing} shows a weak correlation between all the timescales and the stellar mass of the progenitor. This correlation can be explained by the hierarchical growth of structures: statistically, massive objects collapse at later times than lower mass objects. More massive galaxies are thus expected to accrete onto the main halo slightly later than less massive ones \citep[e.g.][]{2015ApJ...807...49W}. Additionally, note that there are 5 progenitors with pericenter distance $>100$ kpc, which are not shown in the left and middle figures of Figure \ref{fig:ordering-pericenter}. The origin of these streams with large pericenter distances is expected to be due to interactions with other substructures within the hosts and will be studied in future work.

\subsection{Validating the Local Velocity Dispersion Criterion}
\label{subsec:disp-criterion-check}
The stream formation time for each object can also be reinterpreted as the stream's age: specifically, how long ago each object began being tidally disrupted. This allows us to validate our classification of phase-mixed objects using their local velocity dispersions: since the phase-mixed objects were once stellar streams, we expect these objects to be older, as a population, than phase-coherent stellar streams.

\begin{figure*}
\plotone{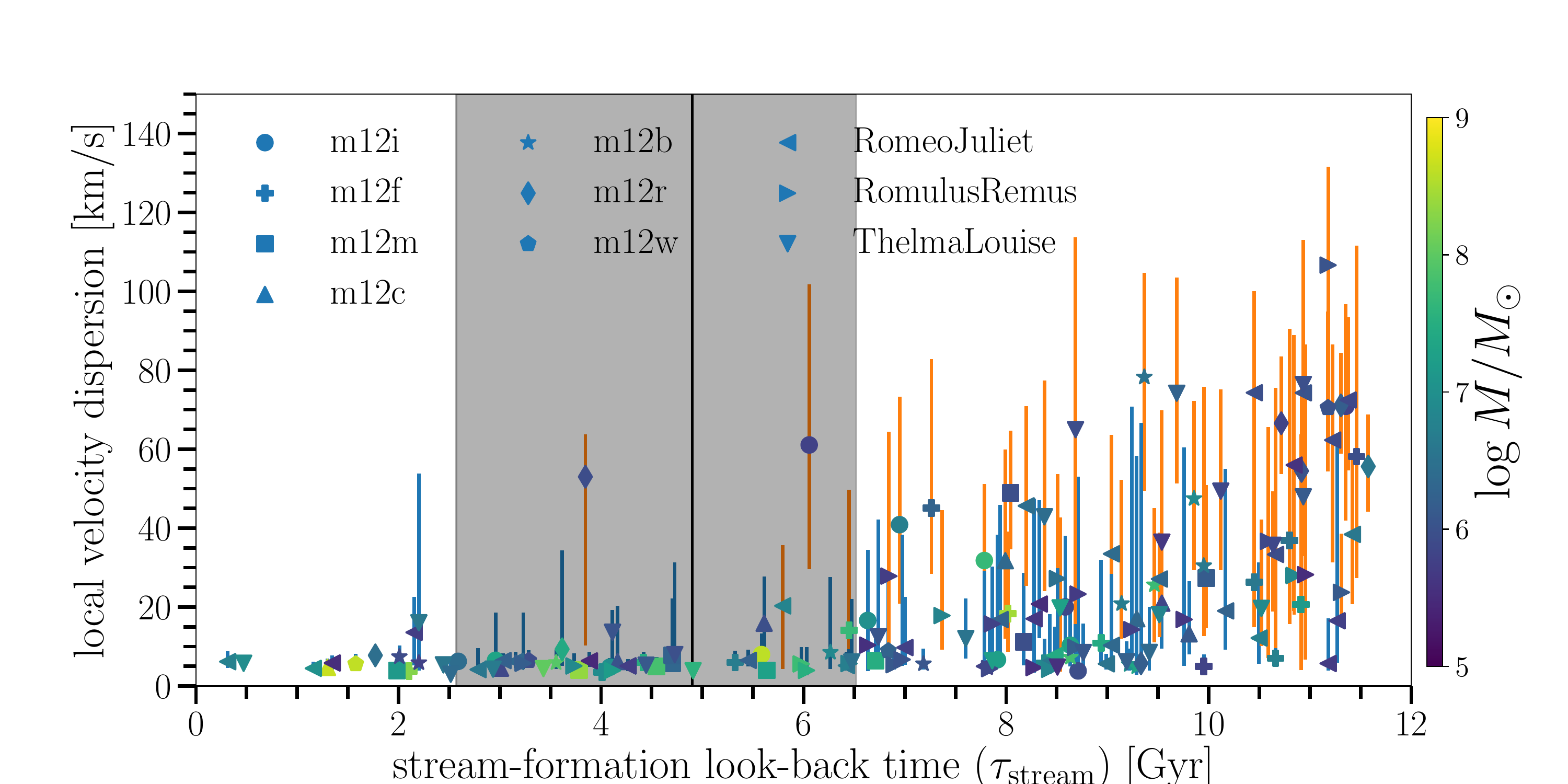}
\caption{Validation of the phase-mixing criterion. Each object's local velocity dispersion $\sigma_\mathrm{local}$ is plotted against its age, which is defined to be its stream-formation time $\tau_{\mathrm{stream}}$ (i.e. time since start of tidal disruption). The marker corresponds to the median local velocity dispersion of star particles in the stream; vertical error bars span 16$^{th}$ to 83$^{rd}$ percentile. The shapes and colors of the markers represent the simulation where each object is identified (see legend) and stellar mass (see colorbar), respectively. Objects that are considered phase-mixed by the local velocity dispersion criterion are shown with orange error bars, while streams that pass the local velocity dispersion criterion are shown with blue error bars. The shaded black region corresponds to the span in the end of the star formation bursty phase/onset of the steady phase in all the hosts. The solid black vertical line at $\tau=4.90$ Gyr is the median of the span. These end of star formation bursty phase times are determined in \citet{2021arXiv210303888Y}. \label{fig:age}}

\end{figure*}

Figure \ref{fig:age} plots each object's present-day local velocity dispersion, with uncertainties corresponding to the 16 and 83 percentile for all particles in the object, as a function of its stream-formation look-back time $\tau_{\mathrm{stream}}$. Each simulation is represented by a unique marker shape; the color of the marker signifies the stellar mass of the object. Objects that are considered phase-mixed by the local velocity dispersion criterion are shown with orange error bars, while streams that pass the local velocity dispersion criterion are shown with blue error bars. The shaded region corresponds to the span in the end of the star formation bursty phase/onset of the steady phase in all the hosts, determined in \citet{2021arXiv210303888Y}, with the median value represented by the black vertical line. In the FIRE simulations, MW-mass galaxies generically have highly time-variable star formation histories at early times but eventually transition to more steady star formation rates after the disk settles \citep[e.g.][]{2015MNRAS.454.2691M, 2017MNRAS.466...88S, 2018MNRAS.473.3717F}. This transition may be due to the virialization of the inner circumgalactic medium, which may stabilize the disk against disruption by stellar feedback \citep[][Gurvich et al. in prep.]{2020arXiv200613976S}.

Phase-mixed objects that are ruled out by the local velocity dispersion criterion are indeed generally older than those that pass the criterion and are classified as streams. The former has a median age of $\sim9.98$ Gyr, while the latter has a median age of $\sim6.45$ Gyr. This validates our use of the local velocity dispersion criterion to separate phase-mixed objects from streams. An additional interesting feature of Figure \ref{fig:age} is the abrupt transition in the number of phase-mixed objects: nearly all began disruption more than 5 Gyr ago.  This is partially a reflection of the time required for an object to phase-mix (further discussion in \S\ref{sec:orbit}) but also reflects the transition between the early epoch of galaxy assembly, which is most chaotic, and the later establishment of a large, stable disk. In this suite of simulations this transition usually occurs at $z\sim 1$ \citep{2018MNRAS.481.4133G}, or about 6.5 Gyr ago. Another contributing factor is the end of the star formation bursty phase, which on average occurs after the disk orientation settlement, at $\sim 5$ Gyr ago \citep[see][]{2021arXiv210303888Y}. Progenitors that fall in and disrupt before this time evolve in a potential with far less symmetry, changing on non-adiabatic timescales, which accelerates the process of phase-mixing substantially relative to the environment after the disk is established. 

\subsection{Progenitors}
\label{sec:proj}
Here we compare properties of stellar stream progenitors with present-day simulated dwarf satellites, and with observational data for satellite galaxies and streams around the MW and M31. We consider three standard relationships: the stellar mass--velocity dispersion relation, the stellar mass--metallicity relation, and tracks in [$\alpha$/Fe]--[Fe/H] chemical abundance space. In all cases we find that the present-day dwarf satellites in our simulations mostly resemble the progenitors of the simulated streams, suggesting that present-day dwarf satellites are useful proxies for stream searches in any of these spaces. We also find that the stellar mass--velocity dispersion relation of our simulated satellites and stream progenitors---and therefore the initial phase-space volume of each simulated stream---is consistent with observations of real dwarf satellites in the MW and M31. This agreement in velocity dispersion relation has been shown for a smaller sample (only satellites in \texttt{m12i}) in \cite{2016ApJ...827L..23W}. Moreover, \cite{2019MNRAS.487.1380G} shows that there is also an agreement between the circular velocity curves of surviving satellites.

\subsubsection{Stellar mass--velocity dispersion relation}
\begin{figure*}
\plotone{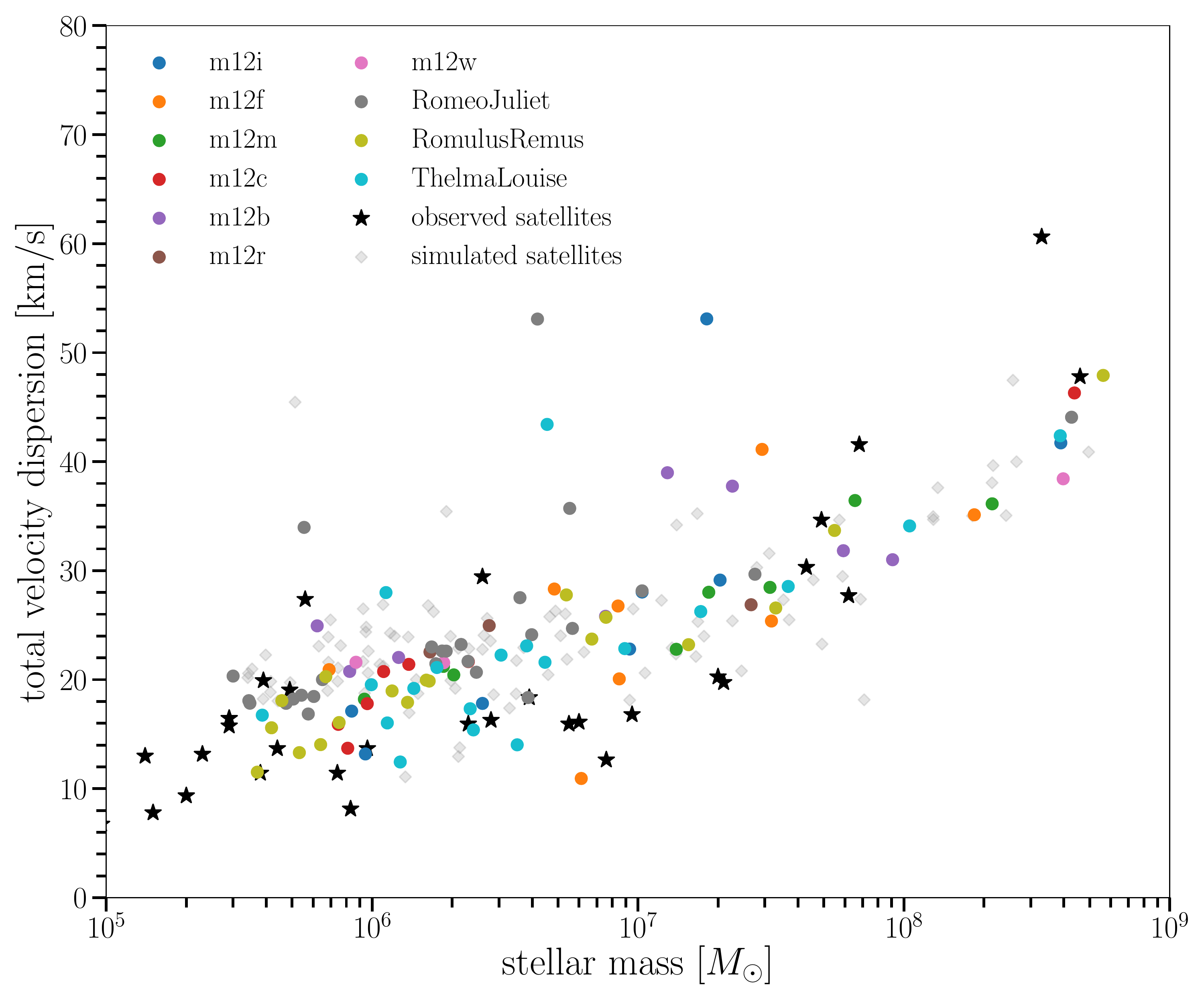}
\caption{Comparison between velocity dispersion of stellar stream progenitors estimated at the peak stellar mass time (dots) and observed MW and M31 satellite galaxies (black stars) as a function of stellar mass. Data for simulated satellite galaxies are also shown (gray diamond). The colors represent individual simulations. Observational data are from \cite{2012AJ....144....4M}. \label{fig:sigma_obs}}
\end{figure*}

The 3-dimensional total velocity dispersions of stellar stream progenitors, simulated satellites and observed satellites are shown in Figure \ref{fig:sigma_obs}. For both stream progenitors and simulated satellites, their total velocity dispersion, $\sigma_\mathrm{tot}$, agree well with the mass--velocity dispersion relation of observed satellite galaxies from \citet[][Table 4; black stars]{2012AJ....144....4M}, with values between $10-30$ km/s for lower mass progenitors ($M_\star < 10^7M_\odot$), and as high as $50$ km/s for higher mass progenitors. \cite{2016ApJ...827L..23W} and \cite{2019MNRAS.487.1380G} present this comparison for surviving satellites and conclude that our simulations do not suffer from too-big-to-fail problem. Our analysis additionally extends this mass--velocity dispersion relation agreement to stream progenitors as well (disrupted satellites).

Compared to stellar stream progenitors, the observed satellite galaxies have slightly lower velocity dispersion for lower mass galaxies ($<10^7M_\odot$), with the values between $5-25$ km/s. This might be due to a selection effect in our sample, as the velocity dispersion is overestimated for objects with fewer number of star particles. However, all observed satellites with velocity dispersion $< 10$ km/s also have stellar mass $< 10^6 M_\odot$. This is very close to the minimum possible stellar mass set by our particle number criterion, which shows our simulations overestimate dynamical masses of systems that have stellar mass $<10^6M_\odot$. In contrast, the velocity dispersions of objects that are $> 10^7M_\odot$ agree well with observations.

\subsubsection{Stellar mass vs metallicity}
\begin{figure*}
\plotone{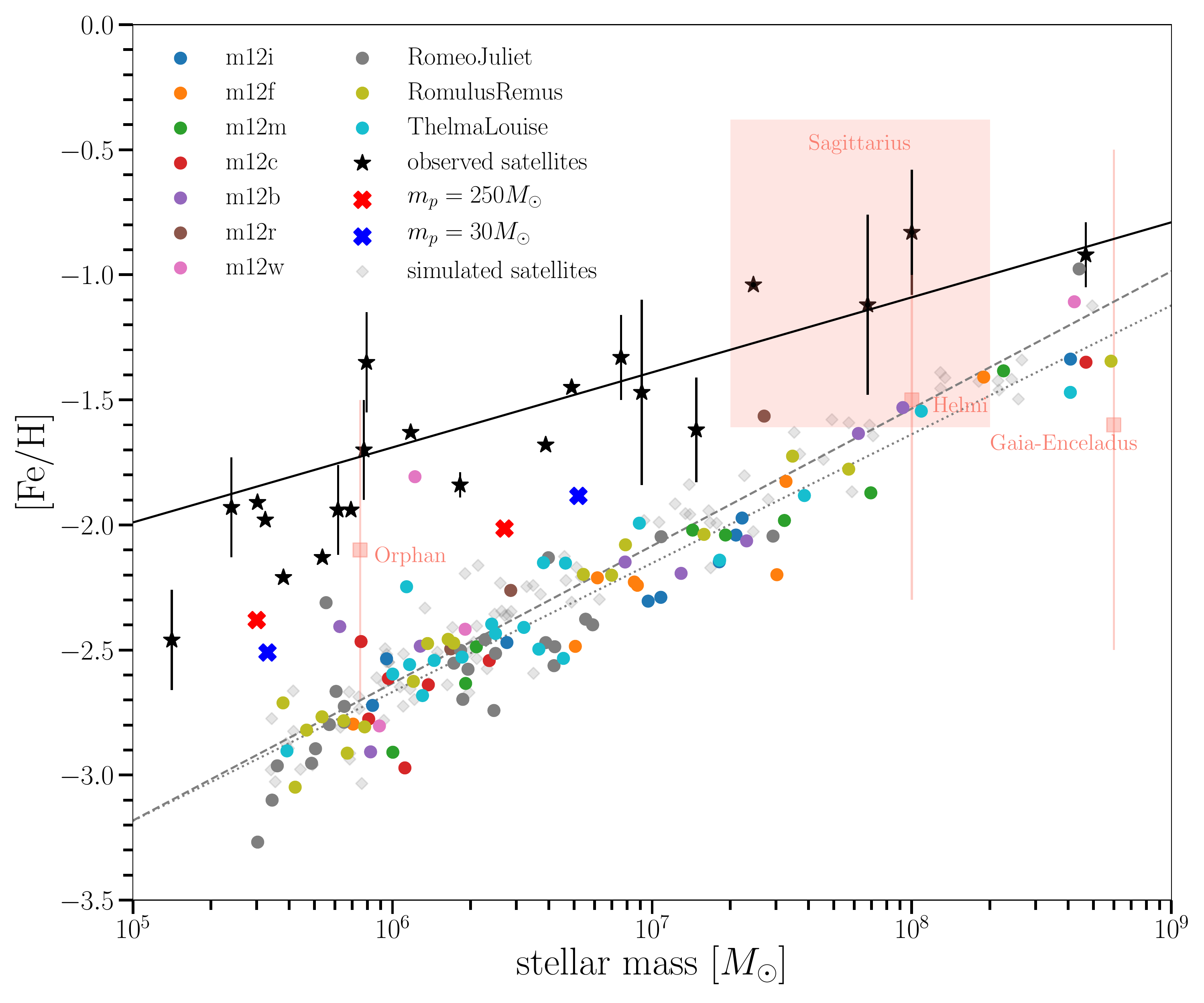}
\caption{Stellar metallicity of stellar stream progenitors labeled by their simulations (colored dots) and present-day simulated satellite galaxies (gray diamonds), compared to observations of satellite galaxies (black stars) from \cite{2013ApJ...779..102K} and streams (red rectangles). The solid line is Equation 4 of \cite{2013ApJ...779..102K}; the dashed and the dotted lines are the least-square fit for the simulated satellites and stream progenitors, respectively. High-resolution simulations of isolated dwarf galaxies from \cite{2019MNRAS.490.4447W} are also shown to approximate the size of resolution effects. These are the red and blue crosses, which are simulations with particle mass resolution of $250\Msun$ and $30\Msun$, respectively. For real streams (red rectangles), the data are from \cite{2017A&A...605A..46M, 2017MNRAS.464..794G} (Sagittarius), \cite{2018Natur.563...85H, 2020MNRAS.493.5195D} (Gaia-Enceladus), \cite{2013ApJ...776...26S, orphan} (Orphan) and \cite{2019A&A...625A...5K} (Helmi).  \label{fig:feh}}
\end{figure*}

Figure \ref{fig:feh} compares the mass-metallicity ([Fe/H]) relations for stellar stream progenitors (dots), present day simulated satellite galaxies (gray diamonds), observed satellite galaxies (black stars) and observed streams (red rectangles). Different colored dots represent streams from different simulations. Observational data is from \cite{2013ApJ...779..102K}. For the simulated systems, we evaluate [Fe/H] for stream progenitors at $\tau_{\mathrm{peak}}$, and for satellites at the present day. The solar abundance is adopted from \cite{2009ARA&A..47..481A}.

For all groups except observed streams, there is a linear relationship between [Fe/H] and $M_\star$, with the more massive objects being more iron-rich. There is no significant difference between stellar stream progenitors from different simulations or between paired and isolated systems. However, there is a significant discrepancy in iron abundance between objects in the simulations and in observations (also noted in \cite{2018MNRAS.474.2194E}). At the same stellar mass, observed satellite galaxies have higher metallicity compared to both stellar stream progenitors and present day simulated satellite galaxies. For the present day simulated satellite galaxies, a linear least-square fit yields
\begin{equation}\label{eq:feh_sim}
    [\textrm{Fe}/\textrm{H}]_{\textrm{sim}} = 0.55\times\log M_\star -5.93,
\end{equation}
where $M_\star$ is the stellar mass. For stream progenitors, the fit is given by 
\begin{equation}\label{eq:feh_st}
    [\textrm{Fe}/\textrm{H}]_{\textrm{stream}} = 0.52\times\log M_\star -5.76.
\end{equation}
The corresponding relationship from \citet[][Equation 4]{2013ApJ...779..102K} for observed satellites is
\begin{equation}
    [\textrm{Fe}/\textrm{H}]_{\textrm{obs}} = 0.3\times\log M_\star -7.69,
\end{equation}
or roughly half the log-slope of the simulated relation; discrepancies are larger for lower mass objects. There are several contributions to the underproduction of metals in these simulations. About 0.3--0.5 dex of the discrepancy appears to be resolution related, as illustrated by the additional points (blue and red crosses) from higher-resolution simulations of isolated dwarf galaxies \citep{2019MNRAS.490.4447W}. The rest of the discrepancy is probably attributable to the supernova delay-time distribution used in the simulations (\cite{2018MNRAS.474.2194E}, Gandhi et al. in prep.), in which the delay before the onset of type Ia supernovae is likely too long and by the metallicity convergence tests in \cite{2018MNRAS.480..800H}.  However, these properties are the same across all our simulations, so we can still draw some conclusions by comparing the \emph{relative} metallicities of different groups of simulated objects. The discrepancies between simulations and observations only impact the metallicity normalization, but not the overall shape, spread and intrinsic scatter in the stellar metallicity distribution function \citep{2018MNRAS.474.2194E}. Relative to the present-day simulated satellites, the stream progenitors have very slightly lower [Fe/H], especially at the upper end of the mass range where the discrepancies are more evident, but are roughly consistent at lower masses. The trend can be explained by the fact that massive satellite galaxies are likely to form star particles long after their infalls (\S\ref{subsec:ordering}) and are likely to still be forming star particles at $z=0$. We see the same trend in Figure \ref{fig:metal_track}. The small discrepancies as a function of stellar mass can be quantified by the differences between Equation \ref{eq:feh_sim} and \ref{eq:feh_st}. This suggests that present-day satellite galaxies are decent proxies to use to estimate [Fe/H] of progenitors of coherent streams in MW-mass galaxies within $\sim0.10$ dex for high-mass progenitors ($M_\star \sim 10^9 \Msun$) or within $\sim0.02$ dex for low-mass progenitors ($M_\star \sim 10^5 \Msun$). 

We also show data from a few MW streams with measured iron abundances. These data are not so well constrained, since for most streams we have only a rough estimate of the mass of their progenitor before tidal disruption. Interestingly, these measurements appear to be all over the place compared to the orderly mass-metallicity relation for satellite galaxies.

\subsubsection{[$\alpha$/Fe]--[Fe/H] evolutionary tracks}
\label{subsubsec:alphaFe}
Chemical abundances of different elements give complementary information on a galaxy's evolution, especially if they are produced by different channels. Fe is mostly produced by Type Ia supernovae, on long timescales compared to $\alpha$-elements produced by type II (core-collapse) SNe from massive stars. Hence, a galaxy's relative abundance of these two types of elements evolves with time, with [$\alpha$/Fe] starting above the Solar value and decreasing over time \citep[e.g.][]{2009ARA&A..47..371T}. This standard track in abundance space, and its variation with galaxy mass and duration of star formation, has been proposed as a basis for chemical decomposition of the stellar halo \citep{2015ApJ...802...48L}.

\begin{figure*}
\plotone{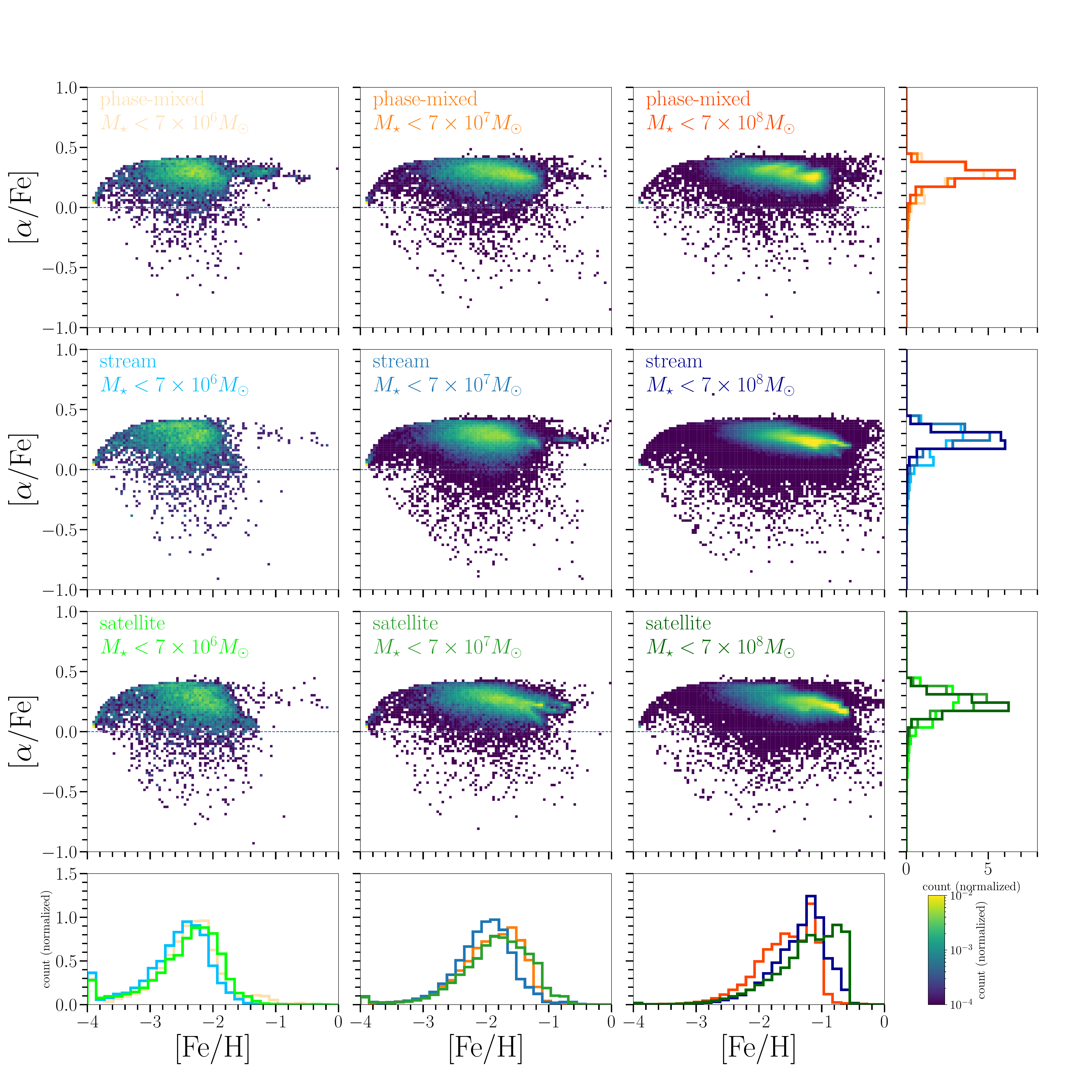}
\caption{Histograms of [$\alpha$/Fe] vs. [Fe/H] and their respective 1D projections for different classes of objects in three stellar mass bins for isolated simulations. Each bin is normalized by the total number of particles in each panel. Each class of objects has a distinct color, with lower mass objects using lighter shades of the color. \label{fig:metal_track}}
\end{figure*}

We study average abundance tracks for the three groups of objects (phase-mixed, streams, and dwarf galaxies) classified using our criteria in $\S$\ref{sec:crit}. The 2D [$\alpha$/Fe] vs. [Fe/H] tracks and their respective 1D projections, stacked for all objects in each panel, are shown in Figure \ref{fig:metal_track}, for objects in different stellar mass ranges. Each 2D histogram is normalized such that the color in each bin represents the number of star particles in the bin normalized by the total number of star particles in that specific class and stellar mass range. For the 1D projections, each class of objects has a distinct color, with lower mass objects using lighter shades of the same color. For phase-mixed objects, the low-mass, intermediate-mass and high-mass groups have the median values of [Fe/H] of -2.36, -1.86 and -1.53, respectively. For streams, the median values of [Fe/H] are -2.49, -2.00 and -1.27. For satellites, the  median values of [Fe/H] are -2.32, -1.80 and -1.16. Similarly, for phase-mixed objects, the low-mass, intermediate-mass and high-mass groups have the median values of [$\alpha$/Fe] of 0.27, 0.27 and 0.23, respectively. For satellites, the median values of [$\alpha$/Fe] are 0.27, 0.27 and 0.25. Lastly, for satellites, the median values of [$\alpha$/Fe] are 0.27, 0.24 and 0.22.

Within the same class of object, the highest stellar mass group is the most iron-enhanced, while the lowest stellar mass group is poor in iron. The [$\alpha$/Fe] distributions are very similar across all stellar mass bins, with the highest stellar mass group having slightly lower [$\alpha$/Fe]. Across different types of objects, the abundance distributions do not depend on mass, except for the high-mass group, in which surviving satellite galaxies have systematically higher [Fe/H] than the disrupted systems. This is in part, because high-mass dwarf galaxies likely have the longest duration of star formation, since they are least likely to be quenched by the present day relative to the progenitors of streams and phase-mixed components (see \S\ref{subsec:ordering}), unlike the smallest dark matter subhalos that can have their star-formation suppressed by a global outside influence, such as the reionization of the universe \citep{2014ApJ...796...91B}.  

In addition to our abundance space analysis, Patel et al. in prep. will examine the origin of the visible features in this abundance space in FIRE dwarf galaxies.

\section{Orbits and alignment with the galactic disk} \label{sec:orbit}
In this section, we study the orbits of stream progenitors and their alignment with the disks of their host galaxies. The orbital plane of each progenitor is characterized by the direction of its total orbital angular momentum at the start of stream formation, while the disk plane is characterized as described in \S \ref{subsec:disks}. We find a slight preference for streams to occupy orbits in the plane of the disk.

\subsection{Disk Angular Momentum Evolution} \label{subsec:disks}

The direction of the total angular momentum of the disk in each simulation is approximated by the direction of the principal axis vector of the moment-of-inertia tensor with the lowest eigenvalue. The principal axes are pre-computed for each snapshot using the youngest 25\% of star particles within the distance that encloses 90\% of the total stellar mass from the center of the host. We compute the disk's total angular momentum explicitly in three of the simulations (\texttt{m12i, m12f, m12m}) to confirm that the principal axes vectors are a good proxy for the disk angular momentum from $\sim 4.6$ Gyr ago onward, where the disks in those simulations are well-established and the direction of their angular momentum is relatively stable.

\subsection{Alignment of stream orbits with the disk plane}

\begin{figure*}
\plotone{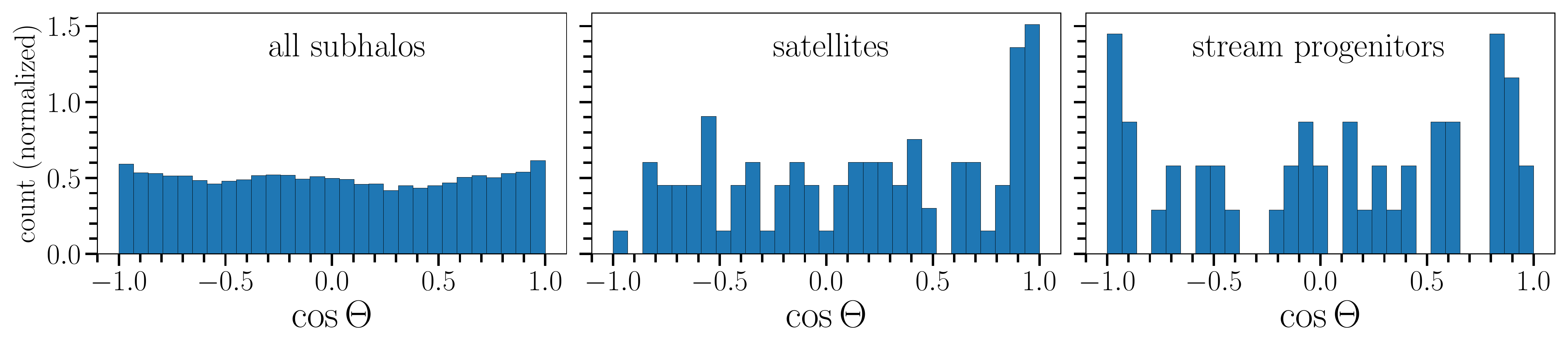}
\caption{Alignment between orbital angular momentum and disk angular momentum for all bound subhalos (dark or luminous) within virial radii of the main host galaxies at present day (left); present-day satellite galaxies with $10^{5}\Msun < M_\star < 10^{9}\Msun$ (middle); and stream progenitors with stream-formation time $\tau_{\mathrm{stream}}<$ 4.6 Gyr (right). The distributions for subhalos and satellites are evaluated at $z=0$, while the distribution for stream progenitors is evaluated at each progenitor's $\tau_{\mathrm{stream}}<$. The angle $\theta$ between the subhalo's orbital angular momentum and the disk's angular momentum is evaluated at present day for the bound subhalos and satellite galaxies, and at $\tau_{\mathrm{stream}}$ for stream progenitors. $\cos\theta=1$ corresponds to prograde orbits in the disk plane, while $\cos\theta=-1$ corresponds to retrogade orbits; $\cos\theta=0$ corresponds to orbits perpendicular to the disk. \label{fig:alignment_all}}
\end{figure*}

We examine the alignment of the total orbital angular momentum of each stream's progenitor with the angular momentum of the disk, both evaluated at $\tau_{\mathrm{stream}}$ to approximate the initial orbits of tidally stripped stars. For this analysis we only include streams whose formation time is less than $\sim 4.6$ Gyr ago in the simulations, for three reasons. First, the principal axis vector with the smallest eigenvalue is a good proxy for the disk's total angular momentum only at later times (see \S\ref{subsec:disks}). Second, the stream-formation time is generally later than both the infall time and the time where the progenitor has peak stellar mass (see Figure \ref{fig:age} and \S\ref{subsec:ordering}). Between infall and stream-formation, there is often enough time for the direction of the disk to settle and the disk to become more stable. Finally, many orbital and dynamical properties of stellar streams, which will be studied later, are determined by the properties of their progenitors at stream-formation time rather than at infall or peak stellar mass time, since the orbit can undergo significant evolution, from dynamical friction and interactions with the disk, in the intervening period.

We compare orbits of bound subhalos (whether luminous or dark) at present day with those of satellites and stream progenitors across all simulations. We select all the bound subhalos at $z=0$ within the virial radius of the main galaxy. For each bound object, we compute its total orbital angular momentum per unit mass with respect to the center of the main galaxy, and then compute the angle $\theta$ between this orbital angular momentum and the disk's angular momentum. Thus, subhalos that orbit in the disk plane have $\cos\theta = \pm1$, while subhalos that have orbits perpendicular to the disk plane have $\cos\theta=0$. 

The left panel of Figure \ref{fig:alignment_all} shows the normalized distribution of $\cos\theta$ for all bound subhalos within the virial radius across all of the simulations (110,094 objects total). The presence of the stellar disk indeed has an impact on the overall orbital distribution of the subhalos: there is a peak in the otherwise almost flat distribution at $\cos\theta \sim \pm1$.  However, this distribution is cumulative across all simulations. The underlying distribution for each individual simulation varies quite significantly from this average profile, which is likely caused by variations in the assembly history of each host galaxy. In some individual cases the majority of subhalos are on retrograde orbits with respect to the galactic disk. To test its similarity with a uniform distribution, we randomly draw 1001 objects ($\sim$10\%) from the distribution in the left panel of Figure \ref{fig:alignment_all} for 10,000 trials. In $55\%$ of the draws, the majority of the objects have $|\cos\theta| > 0.5$. The quantity should be $50\%$ if we draw from a perfectly uniform distribution, indicating a slight preference for subhalos to orbit closer to the disk plane.


Present day satellite galaxies with $10^{5}\Msun < M_\star < 10^{9}\Msun$ and stream progenitors with formation time later than $\sim4.6$ Gyr (middle and right panel of Figure \ref{fig:alignment_all}, respectively) are also slightly more likely to orbit on or close to the disk plane (these are evaluated at present-day for satellites, and at stream-formation time for stream progenitors). However, for satellite galaxies, this is only true for prograde orbits ($\cos \theta = +1$), but not for retrograde orbits ($\cos\theta = -1$). For stream progenitors, there are identifiable peaks in the orbital distribution at $\cos\theta \sim \pm1$, similar to the overall subhalo orbital distribution. Interestingly, the prograde peak is not exactly at $\cos\theta = 1$. For each distribution, we perform random draws of 11 objects ($\sim20\%$ of the sample) 10,000 times. The majority of the objects have $|\cos\theta| > 0.5$ in $68\%$ and $77\%$ of the draws for the satellite galaxies and stream progenitors, respectively. The orbital distributions of these two groups of objects are thus non-uniform, and more skewed towards orbits in the disk plane than for the general subhalo population (see \cite{2020arXiv201008571S} for a detailed analysis of satellite planarity in these simulations).



This finding has several implications. First, streams that orbit near the plane of the disk are more challenging to detect, and this ``selection function'' is likely to limit significantly our count of identified streams detected as overdensities. Streams in prograde orbits are also more likely to be affected by interactions with disk structures like bars, spiral arms, and molecular clouds which can mimic the effects of interactions with subhalos \citep[e.g.][]{2017NatAs...1..633P,2018arXiv180909640B}. Next, it implies that resonant interactions with the disk are slightly preferred, supporting interpretations of structures such as the ``phase-space spiral'' and related disk asymmetries \citep{2012ApJ...750L..41W, 2013MNRAS.436..101W, 2013ApJ...777L...5C, 2018Natur.561..360A} as products of such interactions \citep{2018Natur.561..360A, 2018MNRAS.473.1218L,2018MNRAS.481..286L,2019MNRAS.483.1427L,2019MNRAS.485.3134L}. Finally, the fact that the majority of the satellites are in prograde orbits underlines fluctuations in the orientation of the disk. In our simulations, big mergers can spark the formation of and/or shape the orientation of the stellar disk \citep{2021arXiv210203369S}. \cite{2009Natur.460..605D} argues that resonant interactions on orbits prograde to the disk are most efficient at disrupting these satellites. If the orientation of the disk were fixed, we would expect suppression in the number of prograde satellites, opposite from what we see.


\section{Mass, orbital circularity, and phase-mixing time}
\label{subsec:circularity}
Orbital circularity (the fraction of maximum allowed orbital angular momentum at a given energy) is one of the factors correlated with a stream's mixing time $\Delta \tau_{\mathrm{mix}}$, over which its initially gravitationally-bound stars become mixed with the equilibrium population of stars in the host galaxy. For constant orbital energy, a progenitor galaxy in a radial orbit (low circularity) experiences a steeper potential gradient along its motion; hence, positions of its stars in \textit{phase-space} evolve more quickly compared to a progenitor in a more circular orbit. The increase in stream angular length per orbit can be approximated as \citep[][]{2011ApJ...731...58Y},
\begin{equation}
    \Delta \Psi = \epsilon \left[ \frac{2\pi}{T_\Psi}\frac{dT_\Psi}{dE} \right]_{L=L_{cir}},
\end{equation}
where $L_{cir}$ is the angular momentum of a circular orbit of energy $E$ and azimuthal time-period $T_\Psi(E)$ at radius $r_{cir}(E)$. The prefactor $\epsilon \sim \Delta E$ is the spread in energy of the member star particles, which is approximately
\begin{equation}
    \epsilon = \left(\frac{m}{M_p}\right)^{1/3}\frac{GM_p}{r_p},
\end{equation}
where $m$ is the mass of the dwarf galaxy with radius at the pericenter, $r_p$, and the mass of a host halo enclosed by the pericenter, $M_p$.

We consider objects in circular and radial orbits with the same mass $m$ and energy $E$, assuming comparable $T_\Psi$ and $\frac{dT_\Psi}{dE}$. These assumptions are true for the lowest order approximation of orbits around a point-like mass, where the energy determines the semi-major axis, which is then proportional to the azimuthal period of the orbit. The stream growth rate is
\begin{equation}
\label{eq:dphi}
    \frac{\Delta \Psi}{T_\Psi} \sim \frac{M_p^{2/3}}{r_p} \sim \frac{(\log (r_p/r_s))^{2/3}}{r_p},
\end{equation} 
assuming the standard Navarro-Frenk-White (NFW) profile with $r_p \sim r_s$, the scale radius. Since a circular orbit has the largest $r_p$, stellar streams that are in circular orbits should remain coherent longer than those in radial orbits with the same energy, which should lead to a bias toward circular orbits among older coherent streams. For phase-mixed objects, we thus expect a direct positive correlation between orbital circularity and mixing time. Since the energy spread is also a function of progenitor mass, we also expect more massive progenitors to mix faster.

\subsection{Determining Orbital Circularity}
Because the underlying global potential is time-dependent and not perfectly axisymmetric, orbits are not closed and evolve over cosmic time. Measuring the orbital circularity at different times will thus result in different outcomes. We determine orbital circularity of each object at stream-formation time, since tidal stripping ``freezes'' the orbital properties of stars in a progenitor galaxy with respect to their host.  At any given point along the orbit, the orbital circularity $\eta$ is defined as
\begin{equation}
    \eta \equiv \frac{L(E)}{L_{cir}(E)},
\end{equation}
where $L(E)$ and $E$ are the angular momentum per mass and the total energy per mass, respectively, of the progenitor (streams or phase-mixed objects) at $t_{\mathrm{stream}}$. $L_{cir}(E)$ is the angular momentum per mass of a circular orbit with the same energy $E$. Thus $\eta=1$ corresponds to a perfectly circular orbit at the time of measurement, while $\eta=0$ corresponds to a perfectly radial orbit.

The total energy per mass, $E$, of the orbit is the mean kinetic and potential energy over all particles that belong to the object at stream-formation time. The kinetic energy per mass is the mean of the kinetic energy per mass of all star particles belonging to the progenitor. To estimate the potential energy due to the host galaxy without perturbations from large satellites, we use the AGAMA package \citep{2019MNRAS.482.1525V} to model the smooth component. The potential due to dark matter (within 500 kpc of the host) and gas (within 50 kpc of the host) is represented by an symmetric expansion in spherical harmonics up to $\ell=4$, while the potential of the stars and cold gas  within 50 kpc is approximated by an azimuthal harmonic expansion up to $m=4$. The same radial cut offs are used in the paired simulations---since the paired galaxies are separated by $\sim800$ kpc, the approximated potential around each host galaxy is only computed using particles within the same host. In all cases the potential is set to zero at infinity, so that particles with $E>0$ can be considered unbound. We add the AGAMA potential at stream-formation time at the location $(x,y,z)$ of the progenitor to the kinetic energy to get the total energy per mass, $E$.

For the circular orbit, the velocity of the object can be approximated as $v_{cir}^2(r_{cir})=GM(<r_{cir})/r_{cir}$, while the potential energy per mass is from AGAMA evaluated at $x=r_{cir}$, $y=z=0$. To estimate the radius of the circular orbit $r_{cir}(E)$, we compute the total energy of circular orbits every 2-kpc-interval within the virial radius of the host. The radius that yields the closest value to $E$ is appointed as $r_{cir}(E)$.
The angular momentum per mass of the progenitor is given by $L(E) = |\vec{r} \times \vec{v}|$, while $L_{cir}(E)=|r_{cir} \times v_{cir}(r_{cir})|
$.

\begin{figure}[t]
\plotone{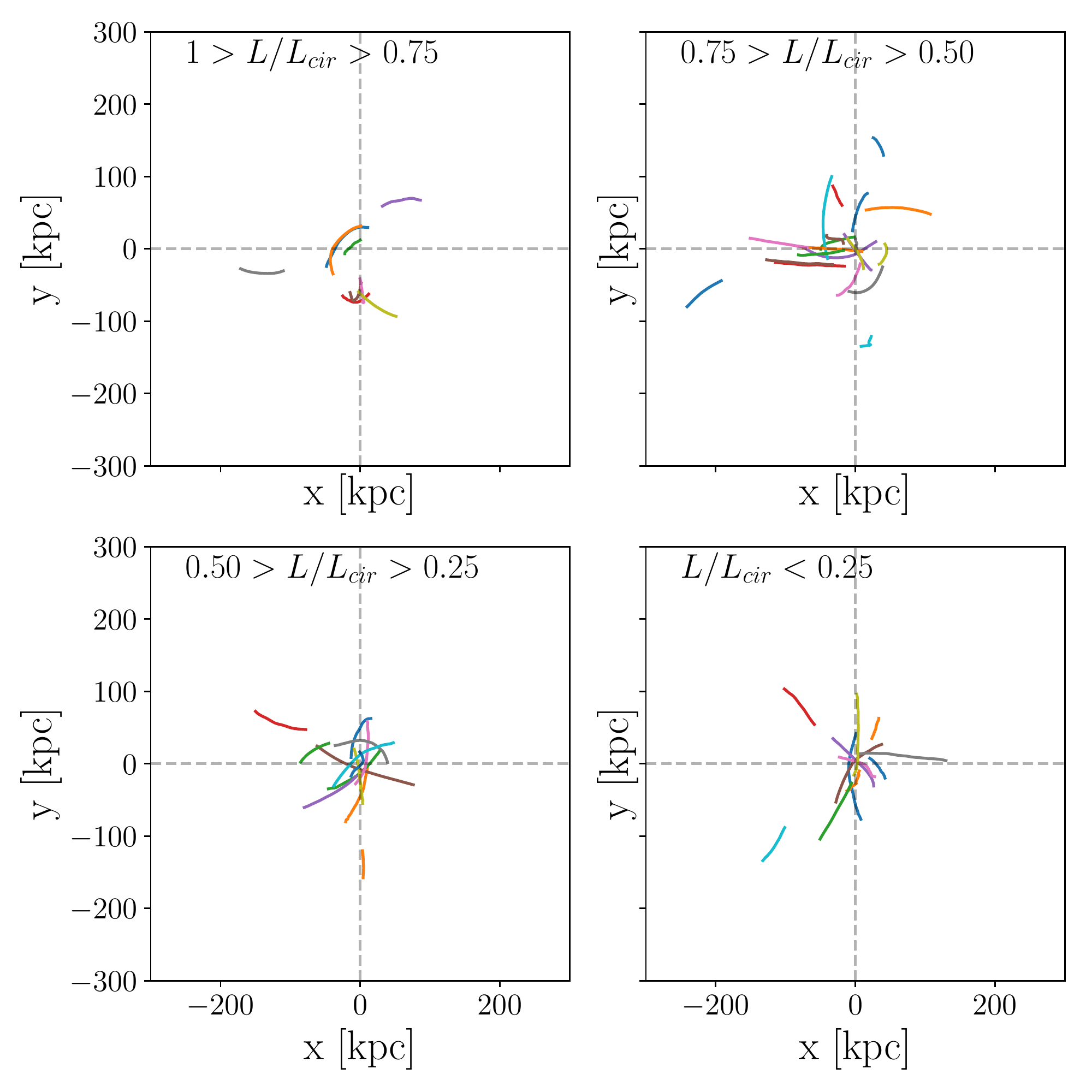}
\caption{Orbital trajectories of stream progenitors and phase-mixed object progenitors in \texttt{m12i}, \texttt{m12f}, \texttt{m12m}, \texttt{m12b} and \texttt{m12r} , spanning $\sim 0.5$ Gyr around their stream-formation time, grouped by orbital circularity $L/L_{cir}$. The trajectories represent motion of the center of the bound part of these objects. The x-y plane here are in the coordinates of the simulations and hence are arbitrary with respect to the host or stream orbit. \label{fig:trajectories}}
\end{figure}

For isolated simulations, $19\%$ have $L/L_{cir}>1$ (indicating a problem), $29\%$ have  $1>L/L_{cir}>0.5$ (relatively circular) and $52\%$ have $L/L_{cir}<0.5$ (relatively radial). For paired simulations, over $60\%$ of the objects have $L/L_{cir}>1$. If the global potential is static and spherically symmetric, the orbital shape, $L(E)/L_{cir}(E)$, should yield a value between 0 and 1 since the circular orbit has maximum angular momentum across all family of orbits with energy $E$. However, the potential in the simulation is time-dependent and not spherically symmetric, especially at earlier times, where the main halo accretes other substructures. The stream-formation time, especially for phase-mixed objects, tends to happen very early on in the simulations. Half of the phase-mixed objects start to form streams over 8.8 Gyr ago (see \S\ref{subsec:disp-criterion-check}). The early stages of evolution in the paired simulations are also much more chaotic, compared to the isolated simulations, and the most-massive and second-most-massive hosts are not well-determined. Thus, the paired simulations and objects with $L/L_{cir}>1$ are excluded in the analysis for the rest of this section. \emph{The large number of objects that must be excluded for having nonsensical circularities underlines the limitations of this simple model for the orbital evolution of streams in realistic cosmological potentials.}

In figure \ref{fig:trajectories}, we show the orbital trajectories of streams and phase-mixed objects in \texttt{m12i}, \texttt{m12f}, \texttt{m12m}, \texttt{m12b} and \texttt{m12r} , spanning $\sim 0.5$ Gyr around their stream-formation time. These objects are grouped based on the circularity, $L/L_{cir}$, of their orbits, and we only consider objects with $L/L_{cir} < 1$. The orbital trajectories of the objects visibly become more radial as the computed circularity $L/L_{cir}$ goes from higher to lower values, confirming that the circularity is at least somewhat correlated with orbital shape.

\subsection{Calculating Mixing Time and Dynamical Time}
\label{subsec:mix_dyn}
In our analysis, the mixing time $\Delta \tau_{\mathrm{mix}}$ is defined to be the difference between the stream formation time and the time when the object is first considered phase-mixed according to Equation \ref{eq:median_vel_dis}. The evolution of the velocity dispersion within each object is not a monotonically increasing function. The period of fluctuations in the local velocity dispersion is equal to the \emph{radial} orbital period, with maximum at pericenter. Hence, the velocity dispersion for some objects crosses the phase mixing criterion threshold multiple times. Our definition of $\Delta \tau_{\mathrm{mix}}$ refers specifically to the difference between the stream formation time and the first up-crossing of the phase mixing criterion threshold. We verified by eye that a subset of objects indeed look phase-mixed after the first up-crossing.


The dynamical time $\Delta \tau_{\mathrm{dyn}}$ relevant for phase-mixing corresponds to the \emph{azimuthal} period, $T_\phi$, of the progenitor around stream formation time (see Equation \ref{eq:dphi}). Rather than computing $\Delta \tau_{\mathrm{dyn}}$ from the approximate potential model, we do so directly from the simulation. We transform the position of a randomly chosen star particle from the simulation Cartesian coordinates into galactrocentric spherical coordinates ($r$, $\theta$, $\phi$) in every snapshot using the instantaneous principal-axis frame determined as described in \S\ref{subsec:disks}, consistent with the assumption of approximate spherical symmetry that underlies the derivation of Equation \ref{eq:dphi}. We begin tracking the particle around 0.5 Gyr before stream formation time and follow the particle until $\phi$ comes back to its starting value.

\subsection{What determines how fast streams phase-mix?}
\begin{figure}[t]
\plotone{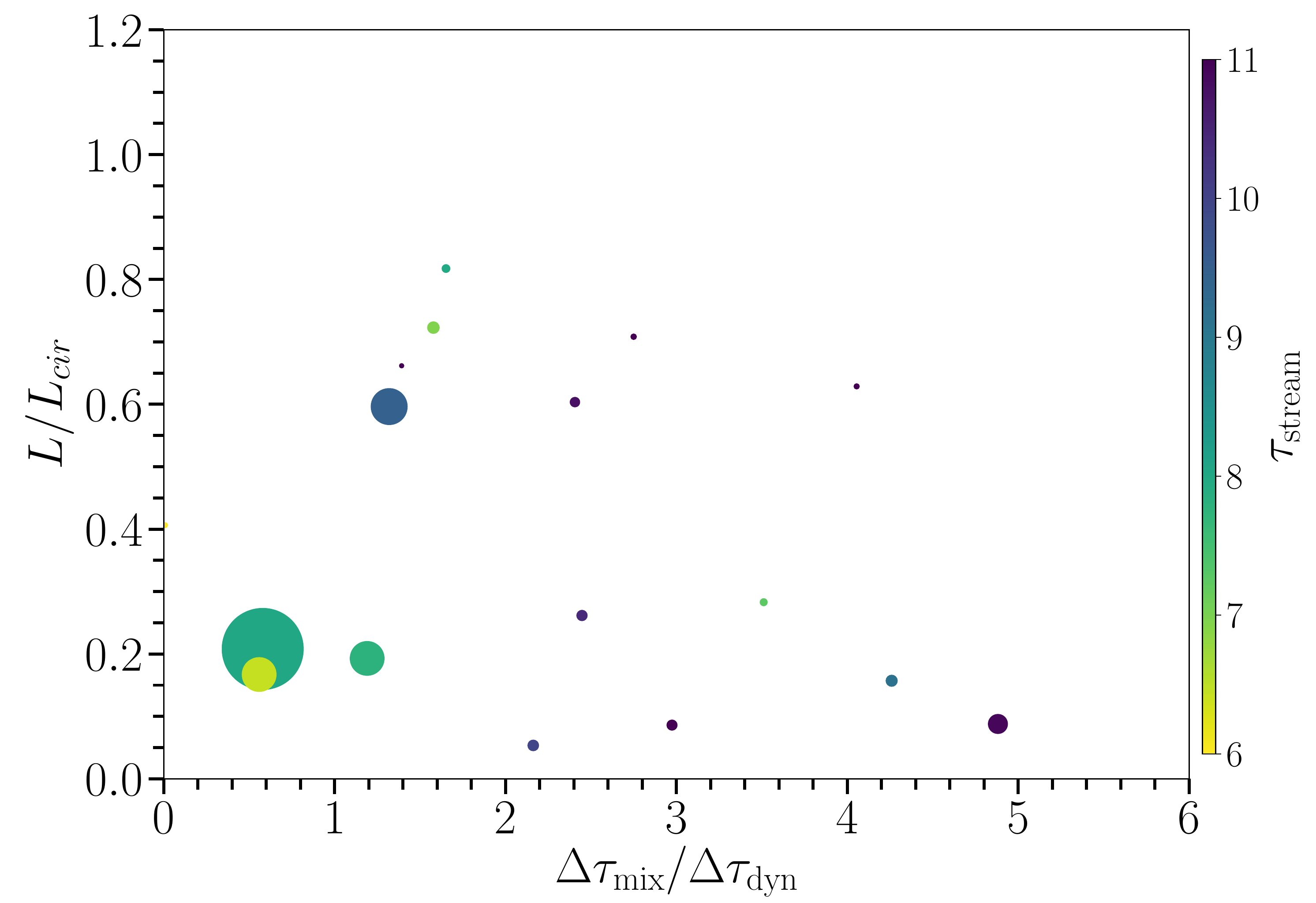}
\caption{Orbital circularity of phase-mixed objects in isolated simulations as a function of the ratio between mixing time and dynamical time, $\Delta \tau_{\mathrm{mix}}/\Delta \tau_{\mathrm{dyn}}$. The stellar mass of each object is represented by the size, while the stream-formation time $\tau_\mathrm{stream}$ is represented by the color of each marker. There is no apparent correlation between the quantities, although no streams on very circular orbits ($L/L_{\mathrm{circ}}\gtrsim 0.5$) mix in less than about 1 dynamical time. Streams that mix the fastest tend to be of higher mass, consistent with Equation \ref{eq:dphi}. \label{fig:mix_cir}}
\end{figure}

Figure \ref{fig:mix_cir} shows the orbital circularity of phase-mixed objects in isolated simulations as a function of the ratio between their mixing time and dynamical time, $\Delta \tau_{\mathrm{mix}}/\Delta \tau_{\mathrm{dyn}}$, described in \S\ref{subsec:mix_dyn}. The color of each point represents the stream-formation time, while the size represents the stellar mass of the object. Contrary to the standard picture described in \S\ref{subsec:circularity}, we do not see a correlation between the orbital circularity with $\Delta \tau_{\mathrm{mix}}/\Delta \tau_{\mathrm{dyn}}$. We argue that this lack of correlation is physical, and not caused by the resolution limitation of our simulations. If the resolution were too low, we would underestimate the mixing time of the objects, especially the low-mass objects with fewer number of star particles. As a result, there would be a trend with stellar mass such that low-mass objects would have the lowest $\Delta \tau_{\mathrm{mix}}/\Delta \tau_{\mathrm{dyn}}$. Instead, we see that high mass objects, for which we can most confidently estimate $\Delta \tau_{\mathrm{mix}}$, have the lowest $\Delta \tau_{\mathrm{mix}}/\Delta \tau_{\mathrm{dyn}}$.

The assumption that the global potential is smooth, static and axisymmetric is clearly not good enough to model the time-evolution of a stream from formation to mixing, especially for objects with very early stream-formation times, when the halo's potential is lumpy and its time-evolution is non-adiabatic. During the early stages of galaxy formation, this chaotic environment accelerates the phase-mixing of accreted satellite galaxies, as illustrated in Figure \ref{fig:age}. \emph{Since the typical time for a galaxy to become a satellite, begin tidally disrupting, and spread out into a stream is long compared to the age of the host galaxy, the non-adiabatic evolution of the galactic potential at early times is potentially one of the main contributors to phase-mixing in the stellar halo.}

\section{Gravitational cooling of streams} \label{sec:dispersion}
This section studies the evolution of the local velocity dispersion as dwarf galaxies evolve into stellar streams. Under an adiabatic potential of the host halo, the conservation of phase-space volume in collisionless systems predicts that streams should grow kinematically colder as they grow in length, albeit in a phase-dependent way \citep{1999MNRAS.307..495H}. The evolution of the local velocity dispersion is thus an important test of whether we are sufficiently resolving our simulated streams. We can also determine to what extent the local velocity dispersion in a stream is a good proxy for the mass of its progenitor. 

\subsection{Comparison of Progenitors' and Streams' Velocity Dispersions}
\begin{figure*}
\plotone{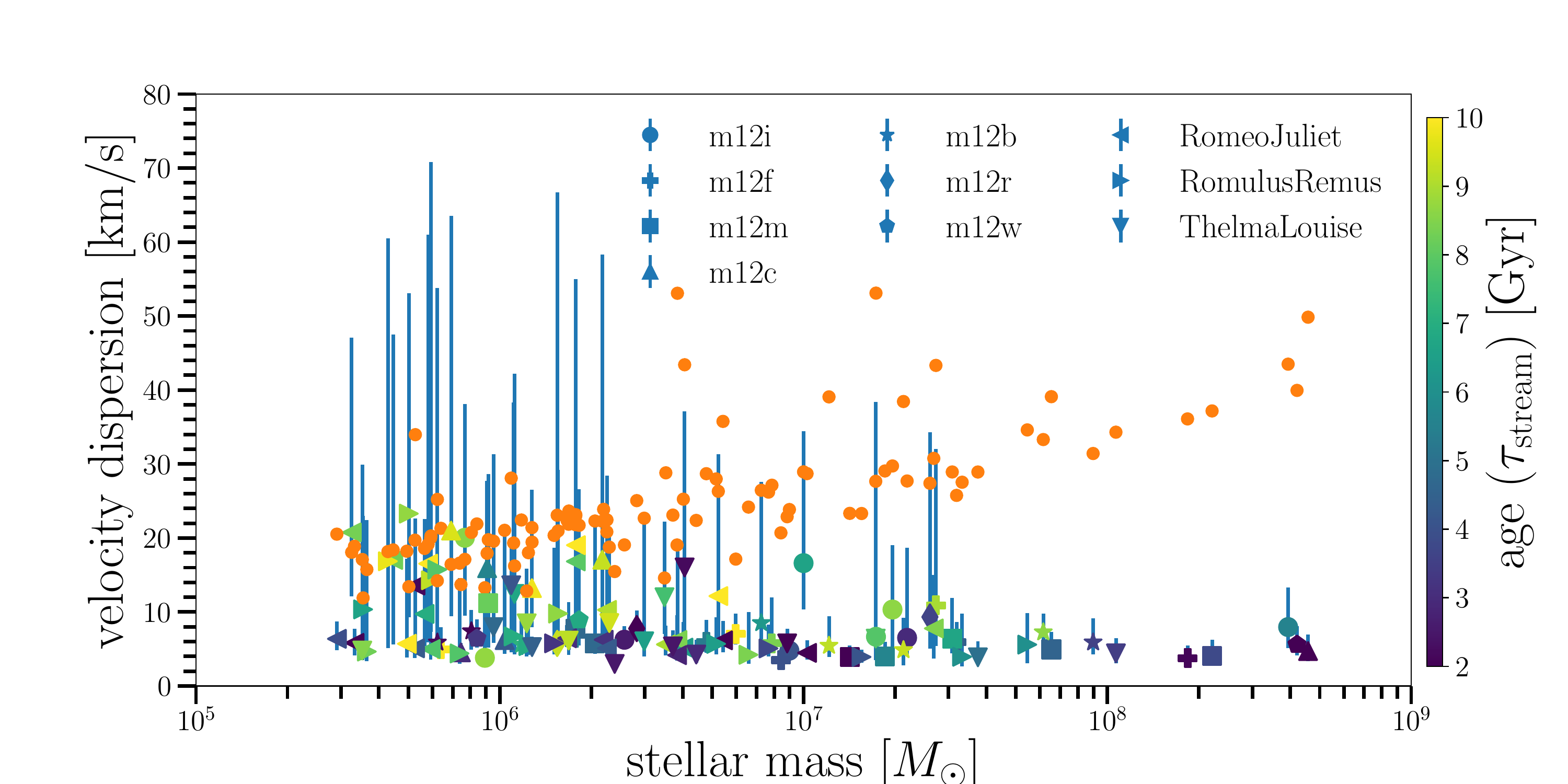}
\caption{Comparison between local velocity dispersions $\sigma_\mathrm{local}$ along stellar streams (blue error bars) and total velocity dispersion $\sigma_\mathrm{total}$ of their progenitors (orange), evaluated at the time when they have maximum stellar mass. The error bars for streams span the $16^{th}$ to $83^{rd}$ percentile of $\sigma_\mathrm{local}$, with the median values shown by the markers. Their colors represent their ages, which are defined to be the stream-formation times $\tau_\mathrm{stream}$ (look-back) described in \S\ref{subsec:infall-form}. \label{fig:vel_dis}}
\end{figure*}

Figure \ref{fig:vel_dis} compares the \emph{global} velocity dispersions of the stellar stream progenitors (standard deviation over all star particles; orange) with the local velocity dispersion when those progenitors become streams (blue). The local velocity dispersions are computed as shown in \S\ref{sec:crit}, where we use 20 nearest neighbors for streams with more than 300 star particles and 7 nearest neighbors otherwise. The median local velocity dispersions along the streams are $\sim 4-20$ km/s, while the global velocity dispersion of the stream progenitors are $\sim 15-50$ km/s. The global velocity dispersion $\sigma$ increases as a function of the total progenitor mass $M$ as predicted for a dispersion-supported system: $\sigma \propto M^{1/3}$. 

The wide range of the local velocity dispersion along a single stream suggests that we should refer to a median value of the local velocity dispersion when we discuss the velocity dispersion of a stellar stream.
Our stellar streams indeed get colder as a function of time, as predicted by the conservation of phase-space volume. 

\subsection{Evolution of local velocity dispersion}
\begin{figure}
\plotone{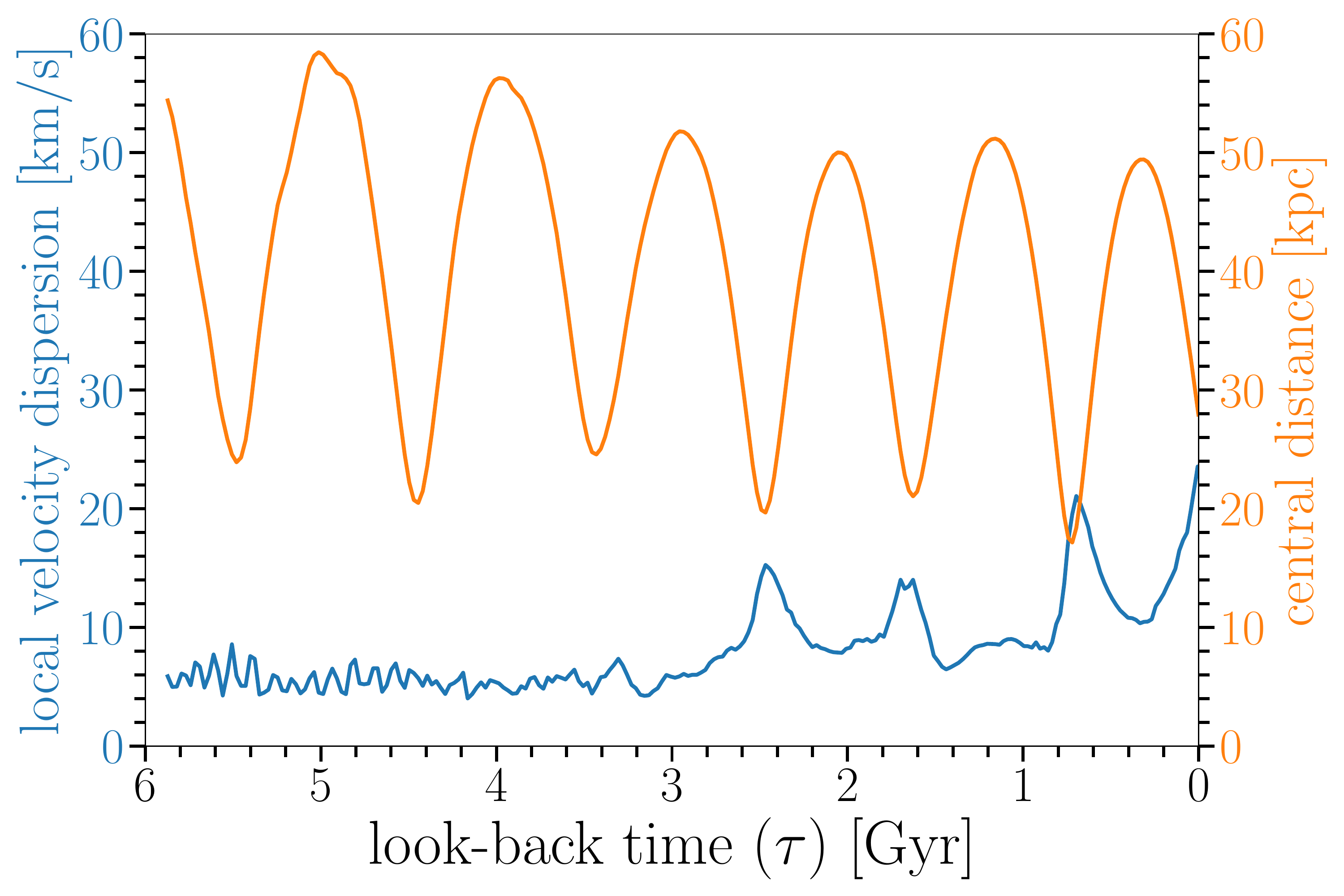}
\caption{Evolution of local velocity dispersion (blue; left y-axis label) and distance to the center of the main galaxy (orange; right y-axis label) as a function of look-back time for one random star particle in a stellar stream selected from \texttt{m12i}. \label{fig:dis_evo}}
\end{figure}

We select a sample stellar stream from $\texttt{m12i}$ to study how its local velocity dispersion evolves with time. This sample stream has about 2000 particles in total, which is roughly the median number of particles for our sample of streams. We track the local velocity dispersion of one randomly-chosen star particle in this stream from the time when the progenitor is still bound until the stream forms. This is shown in Figure \ref{fig:dis_evo} where the blue line represents the local velocity dispersion of this star particle as a function of the simulation time $t$. The start of the simulation corresponds to $t=0$ Gyr, while the present day corresponds to $t\sim 13.7$ Gyr. The orange line represents the distance between this star particle and the center of the main host galaxy. The time elapsed between adjacent pericentric passages, or the radial period $T_r$, is $\sim 1$ Gyr. 

The object has $\tau_{\mathrm{stream}} \approx 8$ Gyr, but this particular star particle leaves the bound part at $t\approx 11$ Gyr (3 Gyr after tidal disruption commences). The local velocity dispersion of this star particle is roughly time-independent before the star particle is stripped from the bound part, even after $t_{\mathrm{stream}}$. After the star particle leaves the bound part ($t > 11$ Gyr), its local velocity dispersion fluctuates between $\sim 5$ and $20$ km/s after the progenitor is tidally disrupted, and is anti-correlated with the star particle's distance to the center of the main galaxy. The orbital phase of the stream governs the velocity dispersion, as the value peaks when the star particle approaches the pericenter of the orbit. This behavior, pointed out in \citet{1999MNRAS.307..495H}, indicates that we are successfully approximating the phase-space evolution of our simulated stellar streams.

It is common to infer the mass of a stellar stream's progenitor by using the velocity dispersion of stars along the stream as a proxy for the dispersion in the progenitor, corrected by a ``gravitational cooling factor'' that scales with the stream age. This example illustrates that this method is complicated by the phase-dependent nature of the velocity dispersion, especially in real observations where we can only observe the brightest fraction of the stellar streams. Fortunately, these brightest parts likely include the remnant of the bound part of the progenitor, which is likely to have a higher stellar density than the stream itself. Using the local velocity dispersion of stars in the bound remnant, in which the fluctuation with orbital phase is small, we can possibly model the mass of the progenitor with higher accuracy. Otherwise, the fluctuation with orbital phase will likely be the dominant source of uncertainty in any mass estimate based on stellar velocity dispersions in streams.

\section{Summary}
In this paper, we present a set of stellar streams, dwarf satellite galaxies and phase-mixed objects selected from the FIRE-2 cosmological hydrodynamical simulations, including 7 isolated MW-mass systems and 3 paired MW-Andromeda-like systems\footnote{full dataset at: \href{https://flathub.flatironinstitute.org/sapfire}{https://flathub.flatironinstitute.org/sapfire}.}. Objects are classified using a set of phase-space criteria (\S\ref{sec:crit}): satellite galaxies have a small size in both position and velocity, streams have a large extent in position and small local extent in velocity, and phase-mixed structures have a large range in both position and velocity. We show that these stellar streams are dynamically cold, and that the median value of the local velocity dispersion within an object, $\left< \sigma \right >$, can be used to separate phase-mixed objects from stellar streams, with only a weak dependence on stellar mass. Applying these criteria, we find a total of 106 simulated coherent stellar streams with stellar masses $10^5 \lesssim M_* \lesssim 10^9 \Msun$ across all 13 MW-mass hosts (\S\ref{table:summary}). We use this sample to study the origin, chemistry, and dynamical properties of simulated stream progenitors. The important findings are the following:\\

\textbf{1. Present-day satellite galaxies are good proxies for stellar stream progenitors} (\S\ref{sec:prop}).  They have similar stellar mass and total mass functions (\S\ref{subsec:massfunc}), implying that satellite galaxies are equally likely to turn into stellar streams independent of mass. The stellar mass--velocity dispersion relation for stream progenitors, evaluated at $t_\mathrm{peak}$, resembles that for real dwarf satellites (\S\ref{sec:proj}). The velocity dispersions of stellar stream progenitors range from 10 km/s for low-mass objects ($M_\star \approx 10^6 \Msun$) to 40 km/s for high-mass objects ($M_\star \approx 10^9 \Msun$). The [Fe/H]--[$\alpha$/Fe] evolution tracks for simulated stream progenitors, dwarf galaxies and phase-mixed objects are similar for objects with $M_\star < 7\times10^7\Msun$, while the high-mass dwarf satellites are slightly more iron-rich compared to stream progenitors and phase-mixed objects with similar masses (\S\ref{subsubsec:alphaFe}). \\
    
\textbf{2. The order in which infall, quenching, and tidal disruption occur for progenitors of stellar streams varies with stellar mass} (\S\ref{subsec:ordering}).  Low-mass progenitors ($M_\star < 2.25\times10^6\Msun$) are likely to have their star formation quenched before their first infall, while most high-mass progenitors ($M_\star > 2.25\times 10^6 \Msun$) have their star formation quenched by the host environment, with many continuing to form star particles and reaching peak stellar mass \emph{after} stream-formation time.\\
    
\textbf{3. The orientation of the galactic disk affects the orbital distributions of all surviving subhalos (luminious or dark), dwarf galaxies and stream progenitors} (\S\ref{sec:orbit}). All substructures slightly prefer orbits that align with the galactic disk plane; the degree of bias is higher for dwarf satellites than subhalos in general, and higher for streams than for dwarf satellites.  Subhalos and stream progenitors appear to equally prefer prograde and retrograde orbits, while dwarf satellites are apparently biased towards prograde orbits.\\ 
    
\textbf{4. For streams that form more than 6--8 Gyr ago, the non-adiabatic evolution of the global potential during the early, chaotic phase of the formation of the host galaxy determines how quickly streams become phase-mixed with the host}, regardless of their orbital circularity or mass (\S\ref{subsec:circularity}). For more than half of the streams in our simulations, a smooth, static, and axisymmetric potential is insufficient to model their time evolution after tidal disruption.\\
    
\textbf{5. Orbital-phase-dependent fluctuations complicate using the velocity dispersion to estimate the mass of a stream progenitor} (\S\ref{sec:dispersion}). Most stream progenitors have a total velocity dispersion $>20$ km/s, with a strong mass-dependence consistent with observations and the theory of dispersion-supported systems. Most of the streams have a median local velocity dispersion $\left< \sigma \right > < 10$ km/s, but this fluctuates by up to a factor 4 with orbital phase (highest at pericenter), which translates to more than a factor 50 in mass. \\
    

The use of cosmological-hydrodynamical simulations allows us to study streams and their progenitors in a realistic system without relying on simplifying theoretical assumptions, and avoiding the strong biases in observational measurements (findings 1, 2 and 3). Simulated streams also serve as a tool to test our theoretical models of streams in realistic MW-mass host environment where many of these assumptions no longer hold exactly (findings 4 and 5). The non-adiabatic time-evolution of the host galaxy (especially in its early stages), the non-smoothness in the global potential and the resolution-dependence in the simulations (see Figures \ref{fig:dispersion_cut} and \ref{fig:feh}) all contribute to the divergence from theoretical predictions and semi-analytic models. However, we have shown that the resolution limit alone is not the major source or attribute to all of differences (see Figures \ref{fig:sigma_obs}, \ref{fig:feh}, \ref{fig:metal_track} and \ref{fig:mix_cir}). These discrepancies are thus mainly physical in nature, and call for more realistic models in the post-\emph{Gaia} era.

\begin{acknowledgments}

NP and RES acknowledge support from NASA grant 19-ATP19-0068. We thank Kathryn Johnston and everyone in the Dynamics group at the Center for Computational Astrophysics, Flatiron Institute, for valuable discussions, suggestions and input.

We also would like to thank the Flatiron Institute Scientific Computing Core for providing computing resources that made this research possible, and especially for their hard work facilitating remote work during the pandemic. Analysis for this paper was carried out on the Flatiron Institute's computing cluster \texttt{rusty}, which is supported by the Simons Foundation and the data release is hosted on Flathub.

Simulations used in this work were run using XSEDE supported by NSF grant ACI-1548562, Blue Waters via allocation PRAC NSF.1713353 supported by the NSF, and NASA HEC Program through the NAS Division at Ames Research Center. 

This research was supported in part at KITP by the Heising-Simons Foundation and the National Science Foundation under Grant No. NSF PHY-1748958. AW received support from NASA through ATP grants 80NSSC18K1097 and 80NSSC20K0513; HST grants GO-14734, AR-15057, AR-15809, and GO-15902 from STScI; a Scialog Award from the Heising-Simons Foundation; and a Hellman Fellowship. CAFG was supported by NSF through grants AST-1715216 and CAREER award AST-1652522; by NASA through grant 17-ATP17-0067; by STScI through grant HST-AR-16124.001-A; and by the Research Corporation for Science Advancement through a Cottrell Scholar Award and a Scialog Award.

\end{acknowledgments}

\software{Astropy (\citealt{astropy:2013}, \citeyear{astropy:2018}), IPython \citep{Perez2007}, Matplotlib \citep{Hunter2007}, Numpy (\citealt{oliphant2006guide}, \citealt{vanderWalt2011}), Pandas \citep{mckinney-proc-scipy-2010}, Scipy \citep{2020SciPy-NMeth}, \texttt{consistent-trees} \citep{2013ApJ...763...18B}, 
\texttt{rockstar} \citep{2013ApJ...762..109B}, \texttt{halo\_analysis} \citep{2020ascl.soft02014W},
\texttt{gizmo\_analysis} \citep{2020ascl.soft02015W}, NASA's Astrophysics Data System.}

\bibliography{references}


\end{document}